# Multi-step deformation experiment and development of a model for the mechanical behavior of polymeric glasses

Grigori A Medvedev*,[1] Enran Xing,[2] Mark D Ediger[2] and James M Caruthers[1],
Purdue University,[1] West Lafayette, Indiana 47907, USA and University of Wisconsin-Madison,[2] Madison, Wisconsin 53706, USA
*medvedev@purdue.edu

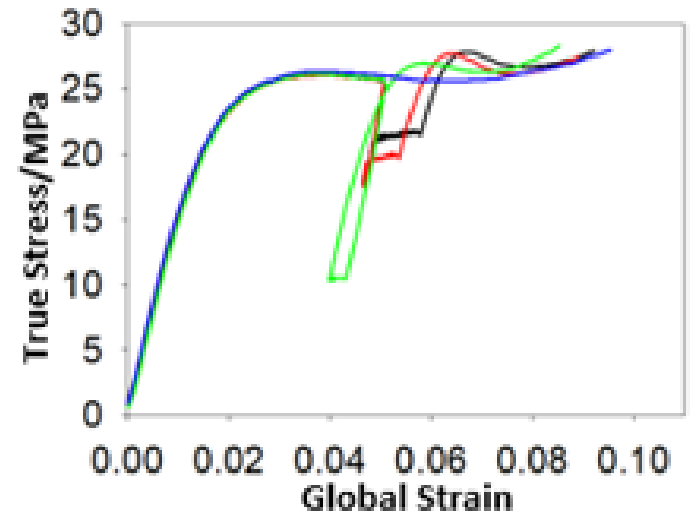

**For Table of Contents use only**

## Abstract

Traditional models for stress-strain behavior of glassy polymers are based on the assumption that the critical features of the stress-strain response can be explained by changes in the molecular mobility. The four-step deformation experiments consisting of (i) an initial constant strain rate loading, (ii) unloading to specified stress, (iii) creep under that stress and (iv) second constant strain rate loading, challenges that assumption. Specifically, existing models fail to predict the experimentally observed large second stress overshoot in case of a slight unloading. Until now there has remained a possibility that the mobility was actually lower in case of a partial rather than complete unloading, which would preserve the main assumption, if not particular details, of these specific constitutive models. By performing direct optical experiments using the photobleaching technique simultaneously with the mechanical four-step experiments it is shown that a lower molecular mobility upon partial unloading does not take place. As traditional models cannot account for these experimental results, a new model has been developed where the changes of molecular structure manifest not in the relaxation time, but in the shear modulus, which is function of an internal variable that is the fraction of the efficiently packed material. This fraction obeys a population balance equation, where the steady-state fraction is controlled by the applied stress. In the absence of deformation, the efficiently packed fraction increases, which explains the increase in the modulus in the course of physical aging below $T_g$. The model qualitatively describes the four-step experiment as well as single step loading experiments.

## Introduction

A key signature of mechanical response of glassy polymers is the stress-strain curve in a constant strain rate deformation as shown schematically as the A-B-C curve in Figure 1. The stress-strain response exhibits a nearly elastic region before reaching yield (i.e., point B) followed by post-yield softening after which the flow stress becomes constant. This stress-strain response is observed for a wide variety of glassy polymers in extension, compression and shear[1-3] as well as for metallic glasses,[4, 5] provided brittle failure is avoided. For polymeric glasses stress-hardening i.e., a post-yield increase in stress, is observed at still larger strains, but hardening behavior is beyond the scope of this paper.

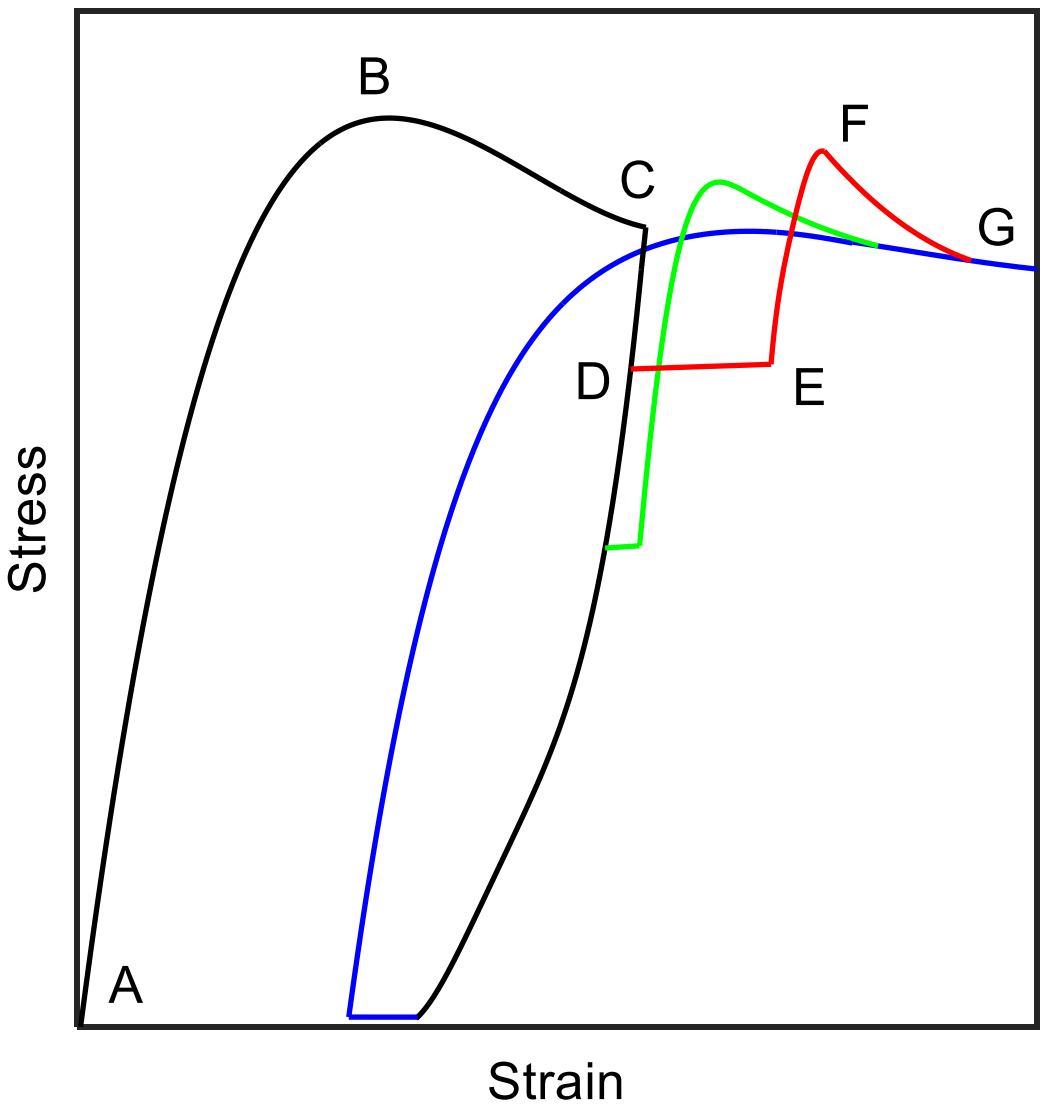

**Figure 1**. Schematics of the stress-strain behavior of glassy polymers in a four-step deformation comprising (i) constant strain rate (CSR) loading: A-B-C, (ii) unloading to specified stress: C-D, (iii) creep under that stress: D-E and (iv) second CSR loading: E-F-G. The strain rate during second CSR is the same as during the first CSR. Colors indicate three unloading/creep stresses: blue – zero i.e., complete unloading, green – intermediate i.e., partial unloading and red – large i.e., slight unloading.

The physical mechanism behind the observed stress-strain behavior has been investigated for over sixty years, where the consensus view attributes the dramatic change from nearly elastic response below yield to the flow response above yield to a change in the molecular mobility, which is effectively defined as the inverse of the current relaxation time. This view arises naturally as many characteristics of glassy behavior are clearly related to mobility. Specifically, vitrification itself is thought to be a manifestation of the slowing down of the relevant molecular motions, so that in the vicinity of the glass transition the average relaxation time of the material exceeds the observation time. From this perspective, the material prior to deformation (i.e., in the state A in Figure 1) has low mobility and the material past yield (i.e., in the states B and C in Figure 1) has high mobility, the latter presumably having resulted from the work of the deformation. In this traditional view, the deformation is roughly equivalent to heating the glass, where yield occurs when the mobility in the material being deformed below $T_g$ is roughly equivalent to the mobility of an undeformed material near $T_g$.

The intuition that there is a dramatic increase in mobility as a glassy material is deformed through yield has been confirmed by direct observation. Ediger and co-workers used a photobleaching technique to optically observe the reorientation of probe molecules during deformation of a polymer matrix.[6] Specifically, it was observed that in a constant strain rate, extensional deformation experiment the average reorientation time of the probe molecules dispersed in poly(methyl methacrylate) and poly(lactic acid) decreased by up to two orders-of-magnitude as the material was brought through yield (i.e., from A to C in Figure 1).[7, 8]

For constant strain rate loading experiments on glassy polymers, the data can be well-described with the basic postulate that the nonlinear viscoelastic behavior is due to an acceleration of the relaxation processes by deformation, where this is the key idea that underlies the existing constitutive models for the nonlinear viscoelastic behavior of polymeric glasses.[9] However, if this postulate truly represents the underlying physics of the deformation process, it should also work for more complex deformation histories. Of particular interest is a four-step experiment, where a polymer in the glassy state is (i) subjected to a constant strain rate deformation, (ii) unloaded to a predetermined stress, (iii) allowed to creep for some period-of-time and (iv) then reloaded at a constant strain rate. This particular multi-step protocol was used by Dreistadt et al for polycarbonate,[10] where the results are schematically shown in Figure 1. Specifically, when the material is unloaded to zero stress, there is no overshoot seen upon reloading; however, if the material is only partially unloaded like from C to D and then allowed to creep, then a second overshoot is observed upon being reloaded at a constant strain rate. As will be described in detail in the next section, a large second overshoot upon partial unloading is inconsistent with the basic postulate that all aspects of the nonlinear mechanical response can be explained by just accelerating the rate of relaxation.

The objective of this communication is to critically study the four-step deformation experiment shown schematically in Figure 1. In the next section a toy model will be constructed that has the essential idea that deformation only accelerates the rate of relaxation. It will be shown that models with this structure are incapable of predicting key features of the second overshoot seen experimentally; specifically, the increase in the magnitude of the second overshoot as the creep stress is increased. Then, an experimental study will be described where both the deformation behavior of a PMMA glass will be measured along with the simultaneous measurement of the molecular mobility using an optical probe. The key result of this experimental study is that the observed change in mobility cannot explain the second stress overshoot – this data unambiguously eliminates the primary postulate that the nonlinear viscoelastic behavior of polymeric glasses is solely due to acceleration of the rate of relaxation by deformation. Then, a second toy model will be introduced where deformation affects the modulus of the material as opposed to just the rate of relaxation, where this toy model qualitatively captures the features of the four-step deformation experiment. Finally, there will be a discussion of the implications of the experimental findings and the new toy model for the constitutive description of the nonlinear deformation of glassy materials.

# Background

<u>Single Step Constant Strain Rate Experiments</u>

The point-of-departure for a description of the deformation behavior of polymers is the well-known Maxwell model, where a one-dimensional toy model will now be used to explain the essential features of the postulate that deformation accelerates the rate of viscoelastic relaxation. The Maxwell model is a linear model given by

$$\frac{d\sigma}{dt} = -\frac{1}{\tau}\sigma + G\frac{d\varepsilon}{dt} \tag{1}$$

where $\sigma$ is the stress, $\varepsilon$ is the strain, and there are two material constants - the relaxation time $\tau$ and the elastic modulus $G$ . Stress and strain are formally tensorial quantities, but the key features of the model are exposed by the one-dimensional scalar form given by eq 1. Under constant strain rate (CSR) deformation, $d\varepsilon/dt$ is constant and eq 1 has a solution for stress that after the initial rise reaches a steady-state value of $\sigma = \tau G \, d\varepsilon/dt$. This looks like experimentally observed yielding with the steady-state stress being the yield stress. However, if the values of $\tau$ and $G$ obtained in small deformation, linear viscoelastic experiments are used, then the predicted "yield stress" is unreasonably high. However, when the actual $\tau$ measured during deformation by the optical probe rotation experiments is used, then eq 1 approximately predicts the correct value of yield stress. Thus, the measured mobility behavior in conjunction with the constitutive model given by eq 1 accounts for single step yielding behavior, where the key nonlinear generalization of the linear Maxwell model is having the relaxation time depend upon the deformation.

An important feature of the stress-strain response for CSR loading is the post-yield stress softening (i.e., the B to C portion of the stress-strain curve in Figure 1), where the causative physical mechanism is the key question. Based on eq 1, one possible scenario would be that upon passing yield point the relaxation time $\tau$ continues to decrease for a while so that the steady-state stress that is eventually reached is lower than the yield stress. However, Bending et al found that within the experimental scatter post-yield softening was not accompanied by noticeable decrease in $\tau$,[7] thereby invalidating the above hypothesis. A key observation about the post-yield softening is that the magnitude of the stress overshoot depends critically on the age of the material prior to deformation, provided all other experimental parameters like temperature, strain rate, etc. are kept constant. Specifically, a material that has been rapidly quenched from above $T_g$ to the temperature at which the mechanical experiment is conducted, exhibits no stress overshoot; alternatively, an aged material exhibits a large stress overshoot, where the magnitude of the overshoot increases roughly as a logarithm of the sub-$T_g$ aging time.[11-15] This "physical aging" is typically described as a relaxation of some structural variable, $S$. After material is cooled into a nonequilibrium glassy state the initial value of the structural variable is $S_{in}$ which is higher than the equilibrium value $S_{eq}$; then, during physical aging under isothermal conditions, the structural variable decreases from $S_{in}$ to $S_{eq}$. The physical nature of the structural variable is still a subject of debate, where several candidates have been proposed such as free volume,[16, 17] fictive temperature,[18] configurational entropy[19] and configurational internal energy.[20]

A candidate physics-based description of the post-yield softening is the combination of the Maxwell model given in eq 1 with an equation for structural relaxation. In the constitutive models that have been developed to describe the stress-softening behavior of glassy polymers,[18-29] it has been postulated that the effect of the deformation induced structural relaxation is to have the relaxation time $\tau$ in eq 1 depend upon $S$. The functional form of the $\tau(S)$ dependence and the equation for the structural relaxation, describing evolution of $S$, vary from one model to another, but the basic idea remains. As a representative example, consider the following equations:

$$\frac{dS}{dt} = -\frac{S - S_{eq}}{\tau} + r\left|\frac{d\varepsilon}{dt}\sigma\right| \tag{2}$$

and

$$\ln\frac{\tau}{\tau_{ref}} = \frac{B}{S} \tag{3}$$

where $\tau_{ref}$ is the relaxation time in a reference state and $r$ in eq 2 and $B$ in eq 3 are parameters. Eqs 1-3 form a complete set, which is solved with initial conditions of $\sigma(t=0) = 0$ and $S(t=0) = S_{in}$. In the absence of deformation i.e., at $d\varepsilon/dt = 0$, structural relaxation/aging takes place, where $S$ decreases toward its equilibrium value, $S_{eq}$. When a large deformation is applied, the decrease in $S$ is arrested and even reversed due to the second term on the RHS of eq 2. This term is proportional to the power (i.e., work per unit time) of the deformation. Eq 2 is similar, but not exactly identical, to the underlying structure of a number of constitutive models; specifically, (1) in the Fielding-Larson-Cates (FLC) model[29] the second term on the RHS of eq 2 is $r\left|d\varepsilon/dt\right|$ and in the Buckley et al model[18] it is $r\left|d\varepsilon_v/dt\right|$ where $\varepsilon_v$ is the viscous strain, (2) in the Chen-Schweizer (CS) model[28] the second term on the RHS of eq 2 is $r\sigma^2/\tau$ and (3) in the Boyce-Argon-Park (BAP)[21] and Eindhoven[13] models the RHS of eq 2 is $(S - S_{eq})\,d\varepsilon_p/dt/\tau$, where $\varepsilon_p$ is the plastic strain. The FLC model does not have a correct small deformation limit, where application of a small amplitude, high frequency sinusoidal strain prevents material from aging, which is not supported by experiments – eq 2 avoids this pitfall. The form of eq 2 for the BAP and Eindhoven models is such that when $S = S_{eq}$ then $dS/dt = 0$ and the structure $S$ remains mired at $S_{eq}$ which causes problems in multi-step experiments – eq 2 does not have that problem, since even if $S = S_{eq}$ $S$ can evolve due to the second term on the RHS. Eqs 1 and 2 are 1st order ODEs, whereas a general nonlinear viscoelastic model derived from the rational thermodynamics' framework[30, 31] is properly described by a set of integral equations, including single-, double-, etc. integral terms.[32] From this perspective eqs 1 and 2 are a special limiting case; however, the predictions of the differential and the more general integral constitutive models are qualitatively similar. Also, a constitutive model must formally use finite stress and strain measures versus the infinitesimal stress and strain used in eqs 1 and 2, where there is some stress-strain nonlinearity due to the finite deformation measures; however, these finite stress/strain effects are small and do not account for the strong nonlinear stress-strain behavior including yield and post-yield softening. The dominant nonlinearity controlling the response in all these models is the strong dependence of the relaxation time on the structural variable, where eqs 1-3 capture the key physical idea that deformation affects the rate of relaxation through an internal structural variable *S* that depends upon the deformation that also evolves with time.

Despite their simplicity, eqs 1-3 qualitatively predict the basic deformation experiments, including the single-step CSR deformation experiment i.e., the A-B-C portion of the stress-strain response in Figure 1. The reason for successful prediction of the post-yield softening is the indirect dependence of the relaxation time on deformation via a differential equation for the internal variable $S$. If dependence of the relaxation time on deformation were via an algebraic equation of the form $\tau = \tau(\sigma)$, there would have been no stress overshoot. The magnitude of the stress overshoot depends on the initial (i.e., prior to deformation) value of the structural variable. Large $S_{in}$ results in no overshoot and small $S_{in}$ that is close to the equilibrium value of $S_{eq}$ results in large overshoot. It should not be concluded that any model having the structure similar to eqs 1-3 will result in successful prediction of the post-yield softening under CSR deformation. The specific form of the equations and the values of the model parameters matter. For example, a much more sophisticated nonlinear thermo-viscoelastic model of Caruthers et al, where the role of the structural variable controlling mobility is played by the configurational internal energy, does not predict the post-yield softening.[33]

The toy model described by eqs 1-3 captures the essence of the state-of-the-art in constitutive modeling of glassy polymers, where much more elaborate models of Anand and collaborators[25, 34] and the stochastic model of Medvedev and Caruthers[27] give qualitatively similar predictions. The shape of the stress-strain curve predicted by the eqs 1-3 model is reasonably close to what is observed in a typical experiment. This is unusual for a single relaxation time model. For example, the original single relaxation time versions of the BAP model,[21] the Eindhoven model,[13] and the Buckley model[18] exhibited yield that was much too abrupt as compared to experimental data, where later versions of these models employed a spectrum of relaxation times rather than a single relaxation time to rectify this problem.[35, 36]

Multi-step Deformation Experiments

The single-step constant strain rate loading experiment shown schematically in Figure 1 (A-B-C curve) is an important, but very simple, deformation history. There have been a few studies where unloading to zero stress has been measured for polymeric glasses,[37-39] but there are very few studies of more complex protocols that have cyclical loading-unloading-reloading deformation steps.[10, 15, 40, 41] Dreistadt et al performed a multi-step experiment shown schematically in Figure 1 (A-B-C-D-E-F-G curve) for polycarbonate in uniaxial compression at 25ºC (i.e., $T_g$-130ºC). Specifically, their deformation protocol consists of the following steps: (i) a constant strain rate (CSR) loading is carried out from A to C, (ii) at C the specimen is partially unloaded so that the stress becomes that of D, (iii) this unloading stress is maintained and the specimen is allowed to creep for a period of time until strain reaches E and (iv) a CSR deformation (with the original strain rate) is resumed and the stress goes from E to F to G. The remarkable feature of the experimentally measured stress response shown in Figure 1 is the 'second stress overshoot' F to G. The second overshoot is observed even when the unloading is slight, i.e., when the creep stress D is only slightly smaller than the flow stress C. In fact, Dreistadt et al reported that the magnitude of the second overshoot *increases* with an increase in the creep stress. This is illustrated by the green curve in Figure 1, where unloading is to a lower stress, and by the blue curve, where unloading is to zero stress (note that during 'creep' at zero stress the specimen begins to recover toward its pre-deformed strain). Recently, we pointed out that the stress-strain behavior observed in this four-step experiment is not predicted even qualitatively by existing constitutive models;[9] specifically, the increase in the second stress overshoot with increase in the creep stress is not predicted.

It is instructive to observe the failure of predicting the four-step experiment by the toy model contained in eqs 1-3. Predictions are shown in Figure 2. The stress-strain curves in Figure 2b show the trend that is exactly opposite to the one observed in the experiments of Dreistadt et al.[10] Increasing the creep stress i.e., the stress to which the material is unloaded following initial CSR loading, results in a decrease and eventual disappearance of the second stress overshoot. The reason for the failure of the model is apparent from the corresponding $S$ vs strain curves shown in Figure 2d. As stated above, the magnitude of the stress overshoot is controlled by the value of the structural variable prior to the CSR ramp; specifically, when the initial $S$ is larger, the overshoot is smaller. This is exactly what is seen in Figure 2. The initial $S$ is the lowest for the first CSR loading and correspondingly the first stress overshoot is the largest. Next lowest initial $S$ is for the case of unloading to zero stress (the blue curve); correspondingly, the subsequent stress overshoot is second largest. Finally, the smallest unloading (the red curve) results in the largest initial $S$, where the subsequent stress overshoot is the smallest.

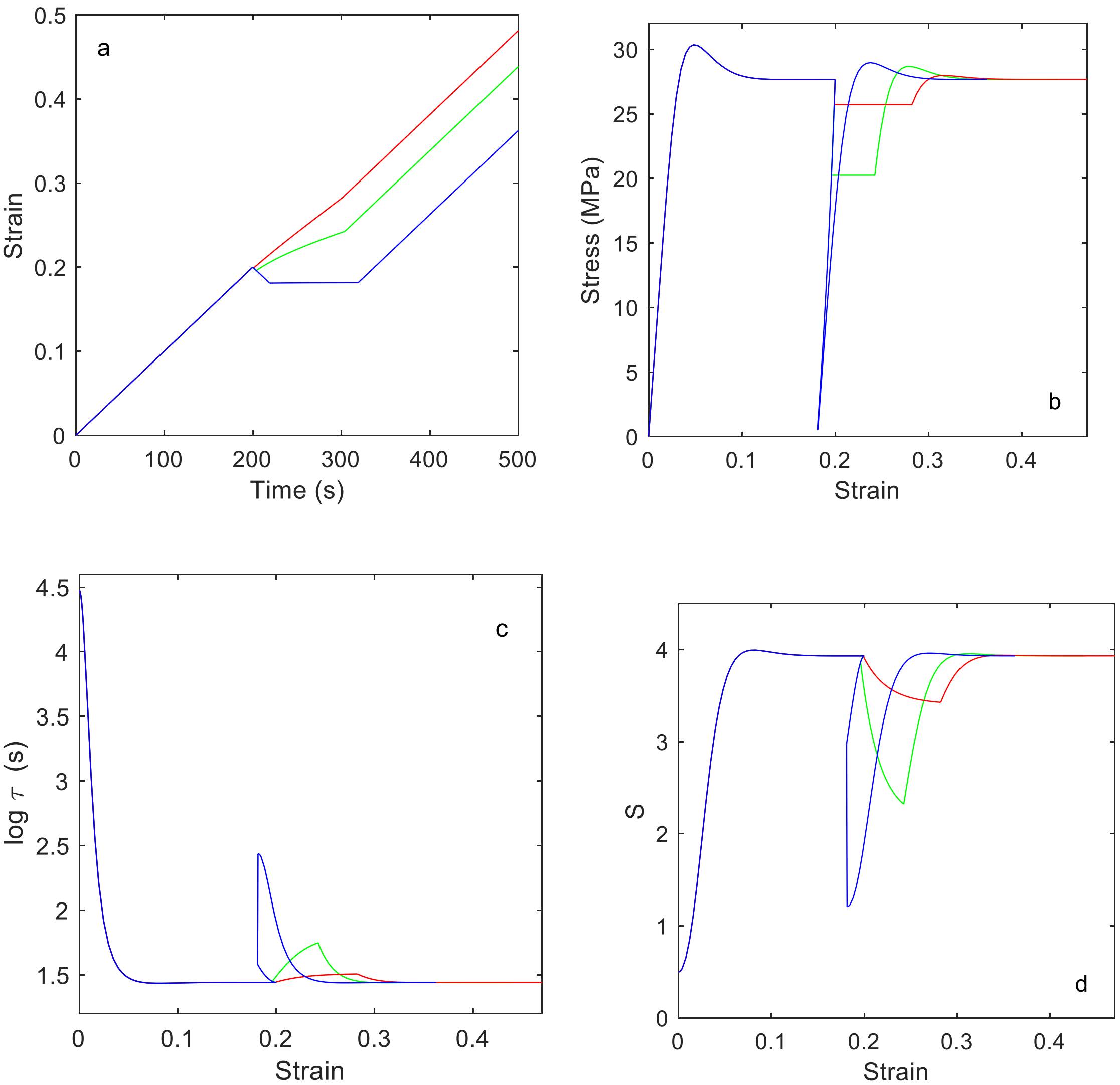

**Figure 2**. The predicted (b) stress, (c) relaxation time $\tau$ and (d) internal variable $S$ vs strain for the deformation histories shown in (a). The internal variable model is defined by eqs 2 and 3 with model parameters: $\tau_{ref} = 10$, $G = 1000$ MPa, $B = 4$, $r = 5$, $S_{eq} = 0.1$ and the initial condition $S(0) = 0.5$; the strain rate is $d\varepsilon/dt = 10^{-3}$ s$^{-1}$.

The monotonic relationship between the deformation and the structural variable is at the heart of the model contained in eqs 1-3. Aging without deformation results in a lower value of the structural variable, where performing mechanical work on the material results in a reversal of the effect of aging, i.e., the larger the work the more "rejuvenated" the material becomes. Upon unloading to zero stress, the aging/structural relaxation resumes; where if the material is then re-loaded the response will be that of a somewhat aged material i.e., it will exhibit a stress overshoot. In contrast, when unloading is only partial, the mechanical work continues to be applied so that material remains "young", where upon re-loading the response will be that of a freshly quenched material and thus will exhibit no overshoot. Details vary, but all existing models, including those much more sophisticated than eqs 1-3, exhibit this basic behavior as has been discussed in detail by Medvedev and Caruthers,[9] where the stress-strain responses shown in Figure 2b are representative of the predictions of these models. It is of course impossible to prove that all possible modifications of a constitutive model where the internal variable just affects the relaxation time will be unable to describe the four-step loading experiment, but the arguments given above clearly show that the second overshoot following a slight unloading does not naturally occur in this class of models.

It seems inescapable that the eqs 2-3 (or the various constitutive models that all have a similar structure[9]) have to be abandoned as they lead straightforwardly to the failed predictions outlined above. But what about eq 1 (or rather its multi-relaxation time generalization) which is known to be successful in describing linear viscoelasticity? Perhaps it can still serve as a basis for describing large deformations provided a different model for structural relaxation than the one contained in eqs 2-3 is developed. The first step in developing a new model is to pose a question: what happens to the mobility (i.e., the relaxation time $\tau$ ) during the course of the four-step deformation experiment? It is well-known that the magnitude of the first stress overshoot is larger in an aged material, i.e., when $\tau$ prior to the first CSR loading (i.e., point A in Figure 1) is much greater than $\tau$ in the flow regime (i.e., point C). Thus, it is reasonable to expect that for the larger second stress overshoot to occur, the $\tau$ prior to the second CSR loading (i.e., point E in Figure 1) should be much greater than $\tau$ in the flow regime (i.e., point G). According to this logic, since the second stress overshoot is larger when the second CSR ramp follows slight unloading (the red curve in Figure 1) and smaller when the second CSR ramp follows unloading to zero stress (the blue curve in Figure 1), one might expect somewhat counter-intuitively that $\tau$ during creep under high stress would be larger than $\tau$ during creep under zero stress. In the *experimental* part of this paper, this hypothesis is tested by performing the *in situ* optical monitoring of the molecular mobility during the four-step deformation experiment, where it will be shown by direct observation that the relaxation time $\tau$ during the creep step is ordered in an intuitive way – larger creep stress corresponds to smaller $\tau$ , i.e., the deformation enhances mobility. Thus, the deformation induced change in mobility is not why a slight unloading produces the second stress overshoot while a complete unloading does not. There must be a different physical mechanism that is responsible for the observed nonlinear deformation behavior of glassy polymers, not the effect of the structural relaxation as described by eqs 2-3. Despite this failure, we believe that the idea of the evolution of the internal structure affecting mechanical response is still physically sound. In the *modeling* part of this paper, we propose a second toy model that, while preserving the Maxwell model in eq 1, replaces eqs 2-3 with a different set of equations, where the effect of structure is manifested in the modulus $G$ rather than the relaxation time $\tau$ . It will be shown that this model successfully predicts the behavior observed in the four-step deformation experiment.

# Experimental

## Sample preparation:

Lightly-crosslinked poly(methyl methacrylate) (PMMA) glasses were synthesized from a stock solution of 98.5 wt% methyl methacrylate (MMA), 1.5 wt% ethylene glycol dimethacrylate (EGDMA) as a crosslinker and ~$5\times10^{-6}$M of N,N′-dipentyl-3,4,9,10-perylenedicarboximide (DPPC) as the optical probe. The stock solution was mixed with the initiator (benzoyl peroxide, 0.1 wt %) and pre-polymerized in a water bath at approximately 345 K to reach the desired viscosity, and then was transferred to molds, which are made with two 2×3 in. glass slides separated with 70μm thick aluminum foil spacers and clamped with binder clips. This mold produces polymer films that have a curved thickness profile that is thinnest in the middle and thickest at each end. The clamped samples were kept at 345 K for 24 h under nitrogen. Subsequently, the films were removed from the molds by sonication and "dog bone-shaped" samples were cut with a custom die. Before deformation, the samples were annealed at 415 K for another 24 h, also under nitrogen. More detailed description of the sample preparation method can be found in previous publications.[6, 7]

Two samples from the same batch were used in the current work - one of which generated data for Figures 3 and 4 in the main text and Figures S3 and S4 in the SI the $\beta$ results in SI are also from this sample, while the other sample generated data for Figures S1 and S2. The samples' thicknesses at their thinnest part are 41 μm and 46 μm, respectively. The glass transition temperature $T_g$ of these samples, as determined from the midpoint of the glass transition from the second DSC scan at 10 K/min, is 399±1 K. Before the experiments, the samples were annealed at 415 K for 30 min to reach equilibrium and then cooled at 2K/min to the testing temperature of 375 K. The deformation started after 20 min at the testing temperature.

### Photobleaching technique:

The photobleaching technique measures the reorientation time of the DPPC probe,[6, 7] which has been shown to be closely correlated with the segmental dynamics of the polymer glass matrix.[42, 43] In these experiments, a linearly polarized 532 nm laser beam was used to preferentially photobleach the dye molecules whose transition dipole moments were aligned with the laser light's polarization state. Then the bleached area was exposed to a weak circularly polarized 532 nm laser beam and the fluorescent light from the unbleached probes were collected and separated into two channels, with polarization states parallel and perpendicular to that of the bleaching laser beam. The time-dependent orientational anisotropy of the dye molecule population, *r(t)*, can be calculated based on the fluorescence intensities of these two channels. By fitting the evolution of *r(t)* with Kohlrausch−Williams−Watts (KWW) function: $r(t) = r(0)\exp(-(t/\tau_{1/e})^{\beta})$, the averaged probe reorientation time, $\tau_{1/e}$, is obtained, along with the anisotropy value at time zero, *r(0)*, and the nonexponentiality factor, $\beta$. We refer to $\tau_{1/e}$ as the segmental relaxation time.[6, 8] When fitting the reorientation dynamics of the undeformed sample, the $\beta$ value was fixed at 0.31 to minimize the uncertainty for relaxation time results as was done previously.[42]

### Mechanical control and data collection:

For these experiments, the tensile deformation was controlled by a programable linear actuator while the force was measured by a load cell in between the linear actuator and the sample, which was held by two clips in a temperature-controlled cell. A more detailed description can be found in previous publications.[6, 7]

The optical set-up during uniaxial deformation of the specimen has been previously described.[7, 44] The optical signals characterizing probe reorientation were collected from the strain localization region of the sample, and its location was identified by an initial experiment in which the samples were stretched by ~15%. As in the work of Lee et al,[6] the local deformation behavior was measured in the same region as the optical signals, by creating photobleached lines that are either perpendicular or parallel to the deformation direction (x-axis) on the undeformed sample and taking images of this pattern during the deformation. The local strain was calculated based on the change of distance between lines perpendicular to the deformation direction. To calculate the true stress value, the contraction of the sample can be determined from the distance of the photobleached lines parallel to the deformation direction. The time-dependent cross section area was calculated with the assumption that the contraction is identical along the y and z-axes.

## Results

The segmental relaxation in PMMA glass during the four-step deformation experiments was monitored using the photobleaching technique. The photobleaching method has been previously used in case of a constant strain rate (CSR) deformation, creep deformation, as well as creep loading and then unloading experiments.[6, 7, 42, 45] Note that the deformation reported here is uniaxial extension, unlike the previous reports on the four-step protocol, where the deformation was uniaxial compression.[10] Also, the current experiment is carried out on PMMA at a temperature significantly closer to $T_g$ ($T_g$-24 K) than the polycarbonate experiments of Dreistadt et al ($T_g$-127 K).

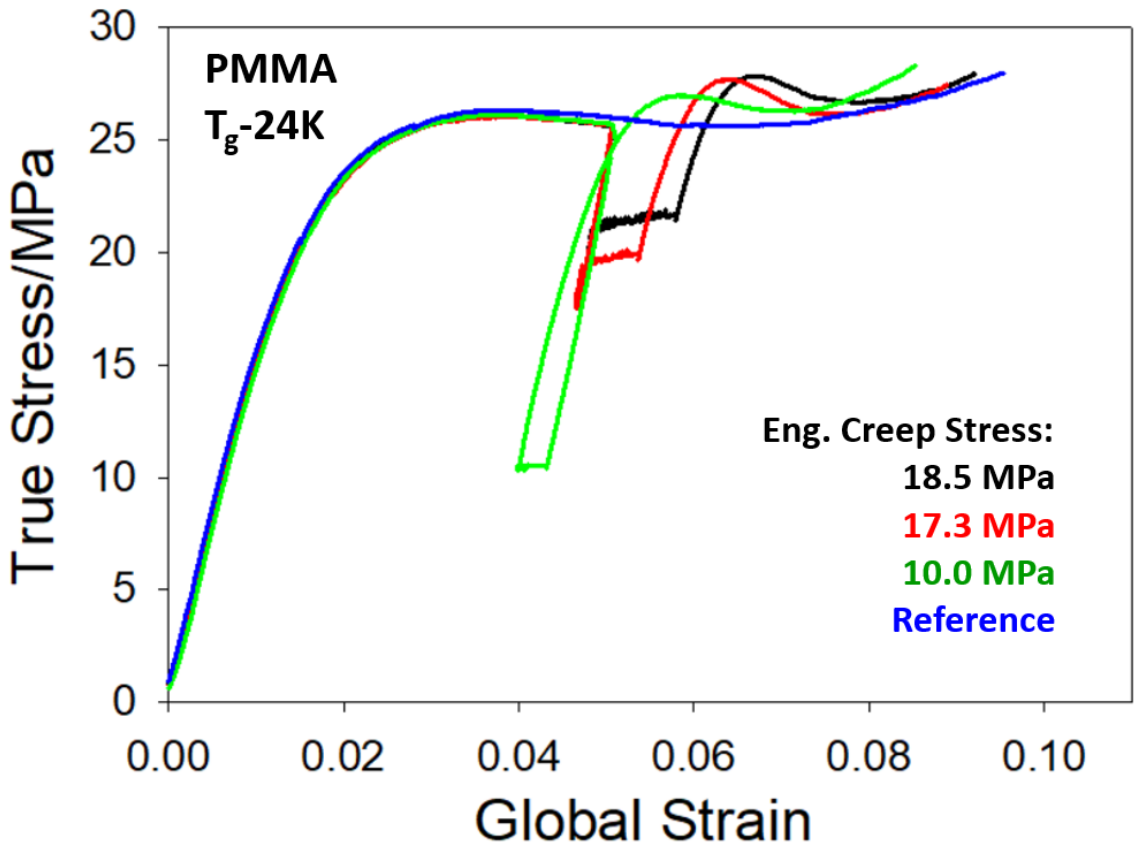

**Figure 3**. True stress plotted against global strain for a PMMA sample deformed using the four-step protocol at $T_g$-24 K, together with a single-step reference experiment. For all the constant strain rate deformations, a global strain rate of $2.1\times10^{-5}$ $s^{-1}$ was used. Between two constant strain rate steps, creep deformation with a constant engineering stress was applied for 1500 s. Experiments using three creep stress levels are shown, all of which show substantial second stress overshoots.

Figure 3 shows the stress-strain profiles of experiments, including the single-step CSR experiment, which is used as a reference, and three four-step experiments, where the creep step was at different stress values. In the four-step protocol, the PMMA glass was first deformed beyond yield at a constant global engineering strain rate of $2.1\times10^{-5}$ $s^{-1}$, and then was partially unloaded and allowed to creep at a constant engineering stress for about 1500 s, and then again deformed into the flow state using the same global strain rate as during the first step. During the creep deformation, the sample extended under 18.5 MPa and 17.3 MPa creep stresses but retracted under a 10.0 MPa creep stress. When the constant strain rate deformation resumed, the second stress overshoot appeared for all three creep stresses, and eventually, the results from the experiments with the two higher creep stresses merged with the single-step reference experiment at a global strain larger than 0.08. Note that the sample was only minimally aged prior to the deformation experiment, where the details of the thermal history are described in the Experimental section. As a result, the first CSR loading exhibits virtually no stress softening after yield, which occurs at the strain of approximately 0.04, where the hardening for this material begins at strains larger than 0.08. The absence of the first stress overshoot makes the second stress overshoot more prominent. In qualitative agreement with the results of Dreistadt et al,[10] the magnitude of the second stress overshoot increases with increase in creep stress.

Because of the strain localization during these tensile deformations, the local strain rate was 3 to 4 times higher than the global value during the first yielding process. At the second yield point in the four-step deformations, the local strain rate was higher than the value at the first yield point by a factor of up to 3. To check whether the second stress overshoot was a result of this increase in strain rate, we performed a second set of experiments in which we reduced the global strain rate in the second constant strain rate step to $1.2\times10^{-5}$ $s^{-1}$. As shown in Figure S1, the second stress overshoot still occurs, with a magnitude comparable to that of the first yielding process. This is consistent with the fact that it takes order-of-magnitude changes in the strain rate to significantly affect the yield stress, as compared to the approximately 40% change in the strain rate from $2.1\times10^{-5}$ $s^{-1}$ to $1.2\times10^{-5}$ $s^{-1}$.

The photobleaching technique provides information about the average segmental relaxation time and the width of the relaxation time spectrum during the deformations shown in Figure 3. The instantaneous relaxation time over a 200-500 s window was fit to the empirical KWW function[46, 47] with the two free parameters: the segmental relaxation time $\tau$ and $\beta$, which describes the width of the stretched exponential function. Figure 4 shows the segmental relaxation time $\tau_{1/e}$ measured by the photobleaching technique for the deformations shown in Figure 3.

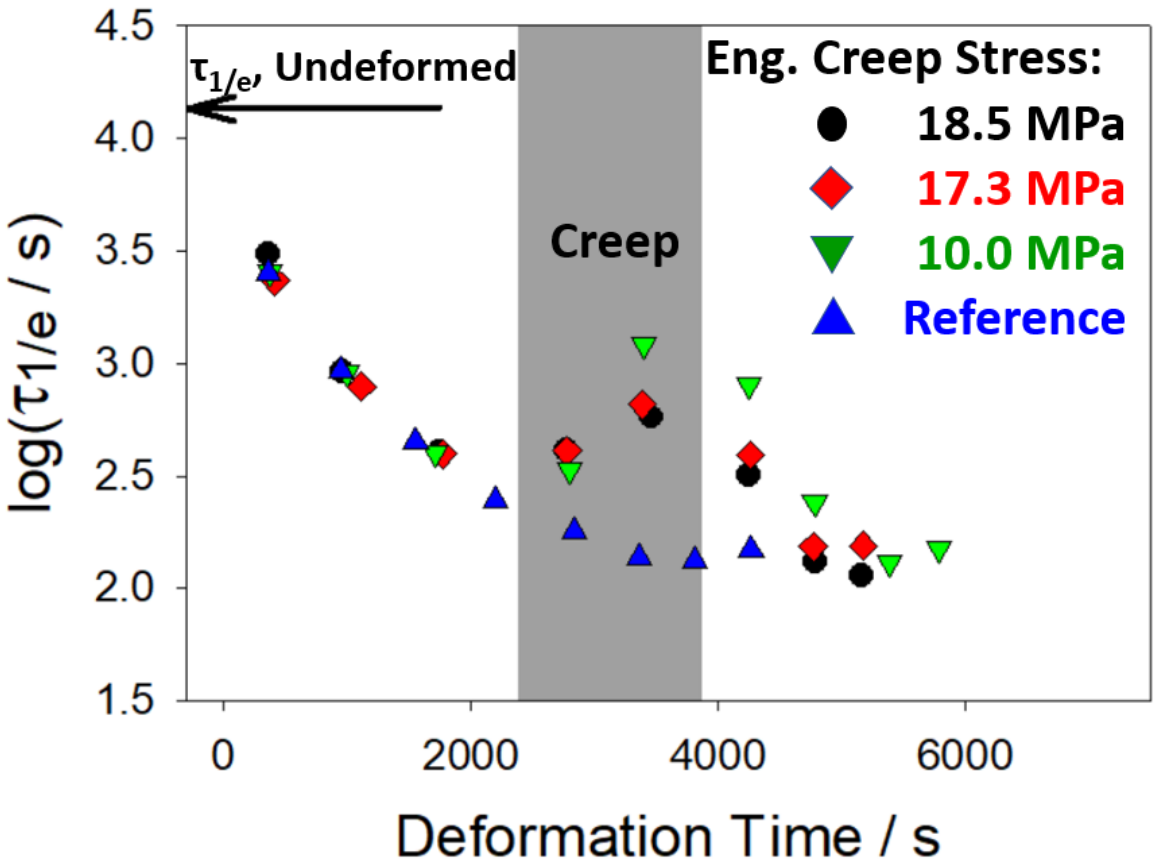

**Figure 4**. Evolution of the segmental mobility as a function of deformation time during the deformations shown in Figure 3. The segmental mobility is represented by the logarithm of $\tau_{1/e}$, which is the averaged segmental relaxation time, on the y-axis. The black arrow indicates the value of $\log(\tau_{1/e})$ before deformation started. And the gray shaded area shows the time period during which creep deformation was performed in the four-step deformation experiments. The acceleration of the segmental mobility during the second constant strain rate step was at least one order of magnitude smaller than that in the first step.

In agreement with previous results,[7] during a single step loading experiment the average segmental relaxation time $\tau_{1/e}$ decreased by about two orders of magnitude from $10^{4.1}$s in the undeformed state to $10^{2.1}$s in the flow state, where it remained constant. In the four-step experiments, during the 1500 s creep step, log $\tau_{1/e}$ increased from the flow state value of 2.1 to a value that depended on the creep stress. These values were 3.1, 2.8, and 2.7 for the creep stresses of 10.0, 17.3, and 18.5 MPa, respectively. This monotonic ordering of the mobility (i.e., the inverse $\tau_{1/e}$) vs. stress agrees with the expectation that upon reduction of the applied deformation the molecular mobility decreases. After resumption of the CSR deformation, log $\tau_{1/e}$ decreased to the flow state value of 2.1. In experiments where following the creep step the stress goes to zero, log $\tau_{1/e}$ gradually increases toward the pre-deformation value of 4.1 (Figures S3 and S4 in the SI). The evolution of width parameter $\beta$ in the KWW function for the deformations shown in Figure 3 is given in the SI, where $\beta$ is approximately constant at 0.6 during the constant strain rate deformation portions of the loading history but decreases during the creep portion of the deformation.

As discussed in the Introduction, in case of a single-step CSR loading there is a correlation between the pre-deformation relaxation time and the magnitude of the stress overshoot; specifically, a larger pre-deformation relaxation time (which is characteristic of an aged material) corresponds to larger stress overshoot. However, examining Figures 3 and 4, a similar correlation does not hold with respect to the second stress overshoot in the four-step experiment. In fact, the exact opposite trend is observed, where the larger relaxation time reached under 10.0 MPa creep corresponds to a smaller second stress overshoot and the smaller relaxation time reached under 18.5 MPa creep corresponds to larger second stress overshoot.

The behavior of the average relaxation time shown in Figure 4 is in qualitative agreement with the predictions of the traditional constitutive models as exemplified by Figure 2c for the model described by eqs 1-3. However, the stress-strain response predicted by the model as shown in Figure 2b is in direct contradiction to the experimentally measured response given in Figure 3. The mechanism that is responsible for the occurrence of the stress overshoot during the second yielding process is not captured by the model, where the key physics is that the primary effect of deformation is to change the relaxation time.

## Toy Model for Deformation of Glasses

In the spirit of the simplified model given by eqs 1-3, we want to develop a second toy model to examine what type of deformation physics is required to describe both the single-step and four-step loading experiments. In developing such a model, we would like to retain the Maxwell element eq 1 for stress. We also believe that the evolution of the structure of the material under the influence of (i) physical aging and (ii) deformation plays a crucial role in the mechanical response. In the previous model the effect of structure was incorporated via the dependence of the relaxation time on the structural variable, but that model has failed to predict the four-step deformation experiment. As an alternative, we propose that the structural dependence enters the eq 1 via the modulus $G$. Specifically, we postulate that a glass forming material exists in two states: efficiently packed and inefficiently packed. Efficiently packed material has a value of the shear modulus $G_1$ and an inefficiently packed material has a modulus $G_2$. By assumption $G_1$ is significantly larger than $G_2$. The new structural variable, $\nu$, is the fraction of the material in an efficiently packed state; correspondingly, the $1-\nu$ fraction is in a state that is packed inefficiently, where $\nu$ is bounded as $0<\nu<1$. At any instant the material has the shear modulus

$$G = \nu G_1 + (1-\nu) G_2 \qquad G_1 >> G_2 \tag{4}$$

As the variable $\nu$ changes with time, so does the shear modulus. Evolution of $\nu$ is governed by a population balance equation, where the efficiently packed fraction grows at the expense of the inefficiently packed fraction and vice versa. Specifically, the efficiently packed material is being 'formed' at a rate, $k_f$, and 'broken' at a rate, $k_b$. Thus,

$$\frac{d\nu}{dt} = k_f (1-\nu) - k_b \nu \tag{5}$$

Eq 5 can be rewritten as

$$\frac{d\nu}{dt} = k_f \left[ (1-\nu) - K\nu \right] \qquad K \equiv \frac{k_b}{k_f} \tag{6}$$

where $K$ is the equilibrium constant. The steady-state fraction $\bar{\nu}$ occurs when in eq 6 $d\nu/dt = 0$; consequently,

$$\bar{\nu} = \frac{1}{1+K} \tag{7}$$

The model parameters $k_f$ and $K$ , and hence the steady-state fraction $\bar{\nu}$ , will in general be functions of both the temperature and the deformation.

The temperature dependent behavior is not the focus of this communication so that the functional form of $k_f(T)$ and $K(T)$ need not be specified. Still, the following assumptions appear reasonable: in a supercooled melt well above $T_g$ inefficient packing is favored, where $K$ will be large, $\bar{\nu}$ vanishingly small and the shear modulus will be on the order of $G_2$ . In a glassy state below $T_g$ efficient packing is favored, where $K$ will be small, $\bar{\nu}$ will be large and the equilibrium shear modulus will thus be large. However, by another assumption both formation and breakage rates are expected to decrease with decrease in temperature so that below $T_g$ the equilibrium $\bar{\nu}$ may not be reached within experimentally accessible times. Still below $T_g$ the modulus described by eq 4 is expected to be on the order of $G_1$ because the fraction $\nu$ , although not as large as the equilibrium value of $\bar{\nu}$ , is significant. During the isothermal physical aging below $T_g$ , the current $\nu$ increases toward $\bar{\nu}$ with increase in aging time. Note that this effect in the evolution of ν is in addition to the evolution of the relaxation time $\tau$ in eq 1 with the aging time.

The key to the model is how $k_f$ and $K$ depend on the deformation. We postulate the following forms:

$$k_f(\sigma) = k_0 \exp\left(-\frac{b}{1+f\sigma^2}\right) \tag{8}$$

$$K(\sigma) = K_0 \exp(c\sigma) \tag{9}$$

where $b$, $f$ and $c$ are model parameters that are all positive. According to eqs 8-9, both the formation rate, $k_f$ , and the breakage rate, $k_b = K k_f$, are accelerated by application of stress, but the breakage rate is accelerated more. As a result, the steady-state fraction $\bar{\nu}$ given by eq 7 decreases under deformation. It is preferable to call it steady-state fraction rather than equilibrium fraction as a material under active deformation is not in equilibrium.

In order to fully specify the model that includes eq 1 for the stress and eq 6 for the internal variable, the deformation dependence of the relaxation time $\tau$ in eq 1 needs to be specified. To better expose the behavior of the model, the preferred choice would be to set $\tau$ to a constant value; in that case the nonlinear effects are solely due to eq 6 i.e., to what the internal variable is doing. On the other hand, if we believe that $\tau$ in the stress equation is the one observed by the optical technique reports, then that $\tau$ is definitely not a constant. We consider two cases:

$$\begin{aligned} &Case\ I: \quad \tau = \tau_0 \exp\left(\frac{b}{1+f\sigma^2}\right) \\ &Case\ II: \quad \tau = \tau_0 \end{aligned} \tag{10}$$

According to eq 10, in Case I the inverse relaxation time has the same dependence on the deformation as the formation rate in the structural variable equation eq 6, and in Case II the relaxation time is a constant. We argue that other possible behaviors of $\tau$ vs deformation, including the one that is experimentally determined, will fall in between these two limiting cases.

The initial condition for stress is straightforward – in the absence of deformation (i.e., $\varepsilon(0)=0$) the stress is zero. The initial condition for the efficient packing fraction depends on the thermal history. When the material is at equilibrium at $T > T_g$, we have $\nu = \bar{\nu}(T)$. As the material is cooled to $T < T_g$, $\nu$ is unable to keep up with increase in $\bar{\nu}(T)$; consequently, $\nu < \bar{\nu}$. To know the exact value of $\nu$, one must describe the cooling process, which in turn requires knowledge of the functional form of $K(T)$. This is beyond the scope of this preliminary study. As a compromise, when modeling a sub-$T_g$ experiment, the efficient packing fraction is assigned an initial value $\nu(0)$ that is lower than the corresponding $\bar{\nu}$, where the latter is given by eq 7. This allows for modeling the effect of physical aging, i.e., the more aged material has larger $\nu(0)$.

Predictions of the four-step experiment using the Case I model are shown in Figure 5, where the strain rate during loading and unloading steps was $10^{-3}$ $s^{-1}$. The values of the model parameters are given in the Table 1. As shown in Figure 5a the model defined in eqs 1, 4, 6 and 8-10 exhibits the main experimentally observed features of the stress-strain data including (i) yield with post-yield softening for the first CSR step, (ii) small overshoot during the second CSR ramp in case of a fully unloaded sample and (iii) larger stress overshoot for the second CSR ramp in case of a partially unloaded sample. The behavior of the inverse formation rate $1/k_f$ and the efficient packing fraction $\nu$ is shown in Figures 5b and 5c, respectively.

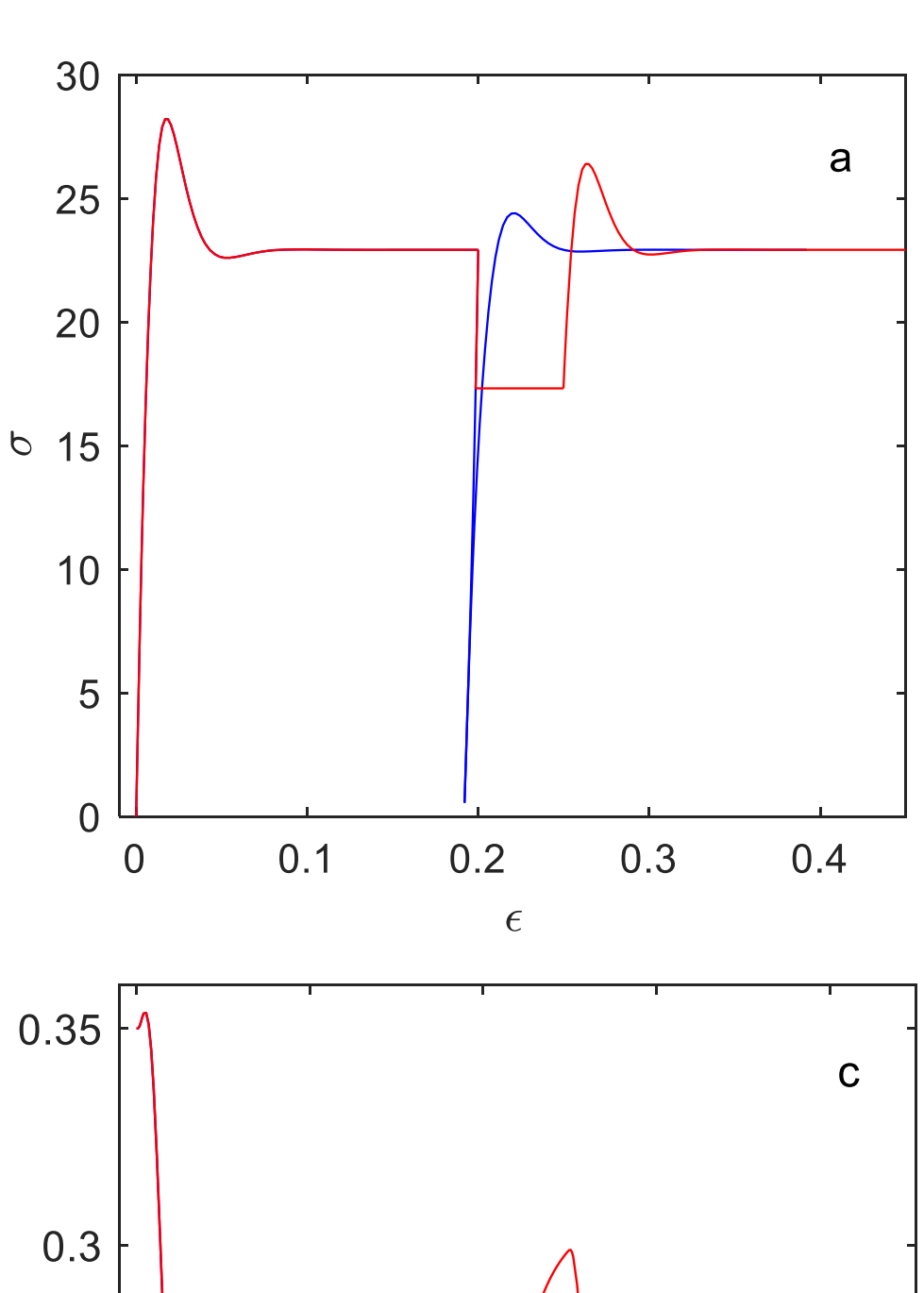

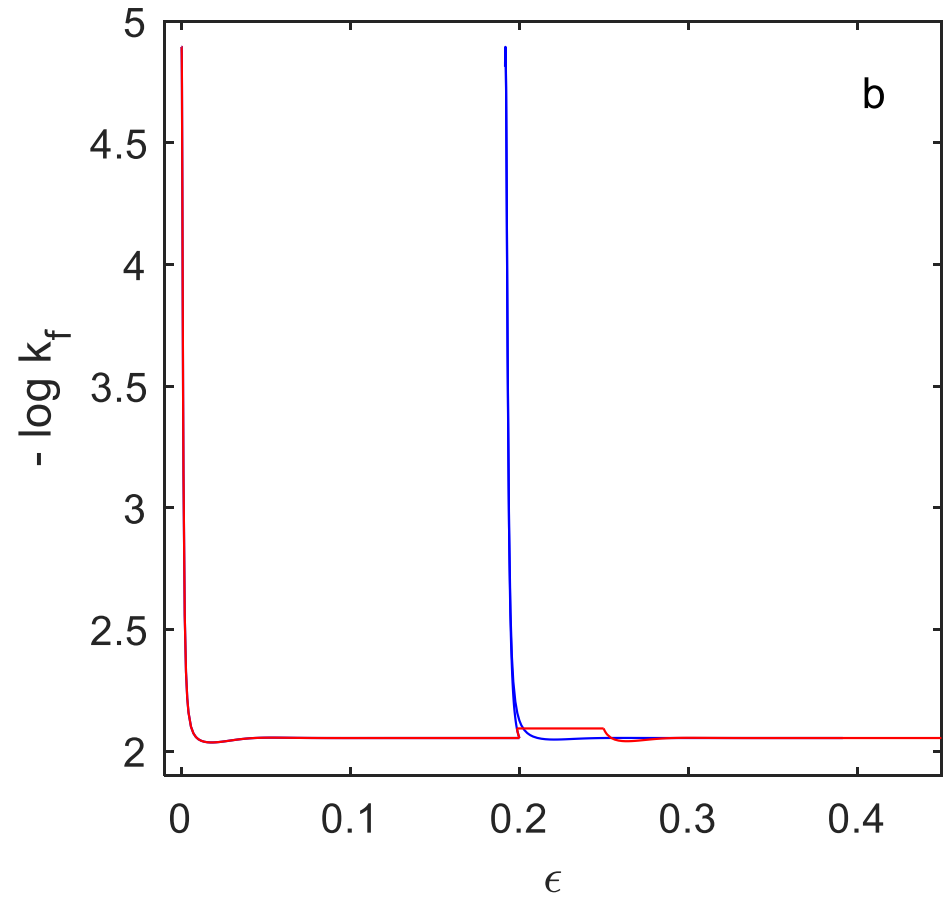

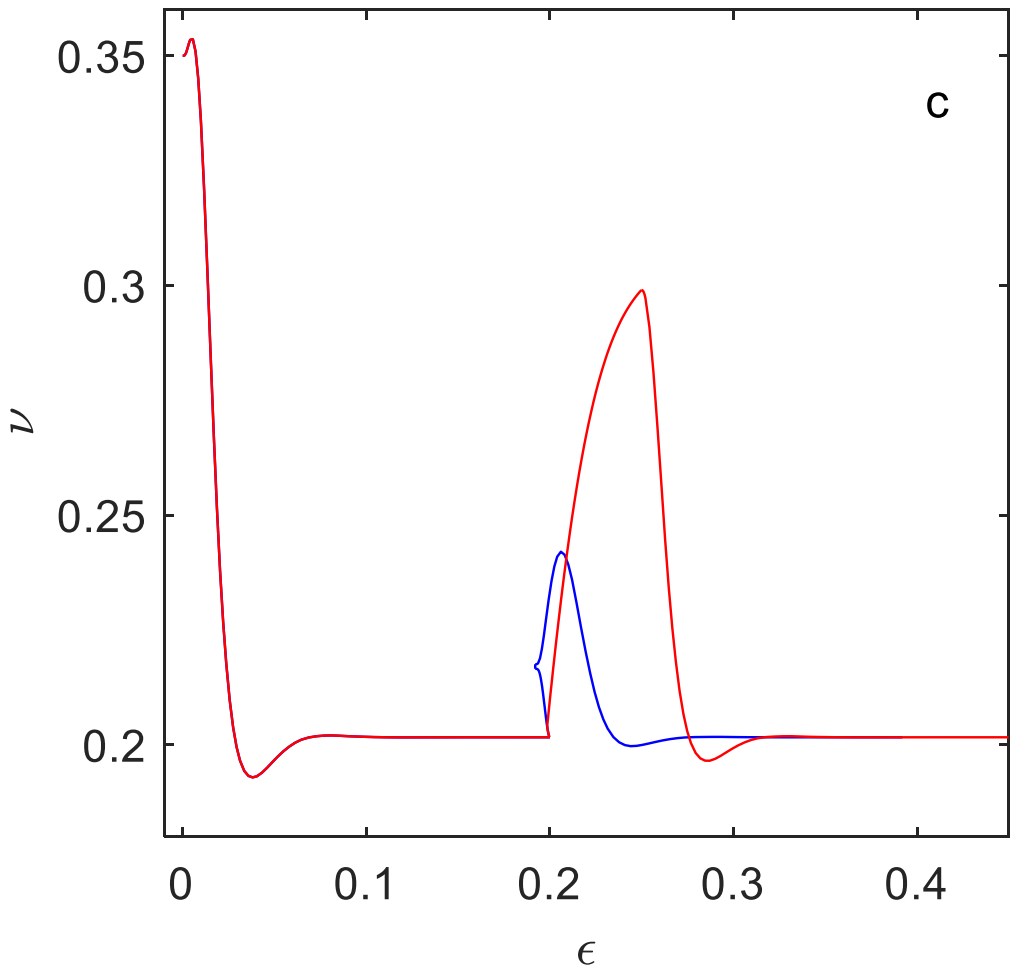

**Figure 5**. Model of the multi-step experiment. (a) Stress, (b) inverse formation rate $k_f$, and (c) efficient packing fraction $\nu$. Unloading to zero stress – blue curve, partial unloading – red curve.

**Table 1.** Values of the parameters for the second toy model.

| $G_1$ | $G_2$ | $\tau_0$ | $k_0$ | $\nu(0)$ | $b$ | $f$ | $K_0$ | $c$ |
|---|---|---|---|---|---|---|---|---|
| $10^4$ | 10 | 10 | 0.01 | 0.35 | 6.3 | 0.1 | 0.4 | 0.1 |

It is instructive to observe which features of the model are responsible for the successful prediction of the four-step experiment. The *first stress overshoot* is a result of the following sequence of events:

1. At $t = 0$ the initial state is $\sigma = 0$ and $\nu = 0.35$. In accordance with Eq 6, $\nu$ wants to relax towards the equilibrium value of $\bar{\nu}_{eq} = 0.71$,but it cannot do so because the formation rate given by eq 8 is too slow. The equilibrium value is dictated by $K$ via eq 7.
2. When deformation is applied the $k_f$ and $k_b$ rate dramatically increase, so (i) $\nu$ is now able to relax and (ii) the value of $\bar{\nu}$ decreases because $K$ increases with deformation in accordance with eq 9. As a result, in its pursuit of $\bar{\nu}$ the variable $\nu$ briefly begins to increase and then sharply reverses course and decreases, as seen in Figure 5c. When the steady-state state is finally reached, we have $\nu = \bar{\nu}_{SS} = 0.2$, where the subscript stands for 'steady-state'.
3. The decrease in $\nu$ causes a decrease in the modulus via eq 4, which results in decrease in the stress i.e., the post-yield softening seen in Figure 5a. Thus, the stress overshoot is predicted.

The height of the overshoot is controlled by two quantities: the initial value of the internal variable $\nu(0)$ and the steady-state value in the flow regime $\bar{\nu}_{SS}$. The larger the difference between $\nu(0)$ and $\bar{\nu}_{SS}$ the larger the stress overshoot, where if there is no difference between $\nu(0)$ and $\bar{\nu}_{SS}$ there is no overshoot. As discussed above, the larger $\nu(0)$ corresponds to longer aging time so the model qualitatively predicts the experimentally observed effect of aging on the yield stress and the associated post-yield stress softening.

The behavior during the *creep* step depends significantly on whether the creep stress is near zero (blue curve) or is 75% of the flow stress (red curve). The duration of creep in both cases is 100 s. Again, there are two determining factors: (i) the drive of $\nu$ toward $\bar{\nu}$ and (ii) the rate of relaxation being high enough (or not) for $\nu$ to move to its target. In case of a creep that follows unloading to zero stress, $\nu$ is frozen at a low value of 0.215 (Figure 5c – blue curve) because the formation rate is slow (Figure 5b – blue curve). In contrast, in case of a creep that follows partial unloading the formation rate is faster (Figure 5b – red curve) and $\nu$ increases almost to $\bar{\nu} = 0.3$ (Figure 5c – red curve). As a result of the previous histories, the *second CSR loading* begins from different values of $\nu$ for the blue and red responses. According to the new model proposed here, a partially unloaded material is akin to an aged material and a fully unloaded material is akin to a quenched material. This is exactly opposite to what happens in the traditional models based solely on the effect of deformation on the relaxation time. That is why the traditional constitutive models for glassy polymers are incapable of predicting the second stress overshoot behavior. Predictions of the four-step experiment using the Case II model (i.e., with the constant $\tau$ ) are virtually indistinguishable from the ones shown in Figure 5 (see SI). This shows that the success of the current model is entirely due to the mechanism contained in the dependence of the modulus $G$ on the efficient packing variable $\nu$ .

*Effect of temperature.* As stated above, although not explicitly modeled here, decreasing the temperature is expected to cause $k_0$ and $K_0$ to decrease as compared to the values given in the Table 1. When smaller values of the parameters are used, the shape of the response shown in Figure 5a is preserved, except the yield stress and the flow stress become larger in agreement with the experimental observations.

*Effect of strain rate*. Increasing the strain rate results in an increase in the steady-state stress, which in turn leads to lower steady-state $\bar{\nu}_{SS}$. As mentioned above, a larger $\nu(0)-\bar{\nu}_{SS}$ difference results in a larger stress overshoot. So, for a constant $\nu(0)$, i.e., a fixed aging time, a lower $\bar{\nu}_{SS}$ means larger stress overshoot. Thus, the experimentally observed increase in stress overshoot with the strain rate is also predicted.

The choice of the model parameters listed in Table 1 is for illustrative purposes, where the predicted behavior is robust with respect to significant changes in the values of parameters. Of course, while the main features of the stress-strain response shown in Figure 5a remain, some finer details of the shape of the curve can be adjusted. For example, a factor of three decrease in the values of the moduli $G_1$ and $G_2$ given in Table 1, while keeping the ratio the same, results in a 35% decrease in the steady-state stress with a corresponding steady-state $\bar{\nu}_{SS}$ of 0.365. Then using the initial value of $\nu(0)=0.35$, no stress overshoot during the first CSR loading is observed. The second stress overshoot is still there, making the stress-strain response similar to the experimental one shown in Figure 3.

## Discussion

The significance of the four-step deformation experiment is that it exposes the inadequacy of the traditional models of the mechanical behavior of glassy polymers that postulate that the primary effect of deformation is a change in a structural variable $S$ that only affects the relaxation time(s). The failure of the models is qualitative, where the predicted trend is exact opposite of what is observed experimentally. Specifically, experiments show that the magnitude of the second stress overshoot, observed after the sample had been partially unloaded and allowed to creep for some time, *increases* with the creep stress. In other words, only slightly unloaded sample exhibits a large second stress overshoot and completely unloaded sample exhibits either a small overshoot or no overshoot at all. The failure to predict this behavior is characteristic of all classes of existing constitutive models, including the viscoplastic models,[18, 21, 23, 25, 34] viscoelastic models[19, 20] and the stochastic constitutive model.[27] Although all these models differ greatly in the mathematical form and complexity, they are based on a single physical idea – that the mechanical behavior is dictated by the molecular mobility; roughly, when mobility is low material behaves elastically and when mobility is high material flows. In the plasticity-based models this idea is expressed in the plastic flow criterion and in the viscoelasticity-based models in the dependence of the relaxation time on deformation. The idea that the mobility is enhanced under active deformation is physically sound and has been confirmed by direct observations via optical means.[6] However, as the failure to describe the four-step experiment proves, this idea alone is insufficient. Here we proposed a different physical mechanism which is based on the postulated effect of deformation on the elastic modulus rather than the relaxation time. A toy model based on this mechanism successfully predicts the four-step experiment.

The traditional constitutive models predict that when the relaxation time just prior to a constant strain rate deformation is larger there will be a larger stress overshoot, where for single-step constant strain rate loading experiments this is in agreement with the optical probe measurements.[7, 8] This postulate was critically examined for the four-step experiment with PMMA, where (i) the magnitude of the stress overshoot increased as the creep stress was decreased in agreement with the four-step experiments of Dreistadt et al[10] but (ii) the molecular mobility at the end of the creep step as measure by optical probe rotation was only slightly larger than the steady-state mobility, which is insufficient to explain the second stress overshoot observed experimentally. This critical experiment conclusively shows that second stress overshoot peak is not controlled by the molecular mobility which is the key postulate in the traditional constitutive models for glassy polymers.

In the previous sections we introduced two toy models: the model described by eqs 1-3 that represents the traditional constitutive models (and although having a simple mathematical form, reproduces the basic features of the more elaborate models) and a new model described by eqs 1,4,6, 8-10. Both models have similar mathematical form being the sets of the 1st order ODEs for two variables, where the first variable is stress and the second variable is a structural descriptor. It is instructive to see why the traditional model fails and the new model succeeds in capturing the trends observed in the four-step deformation experiment. In both models the magnitude of the first, second and any subsequent stress overshoot is controlled by the difference between the value of the structural variable (i.e., $S_{in}$ for the traditional model or $\nu_{in}$ for the new model, respectively) prior to the CSR ramp and the steady-state value of this variable (i.e., $S_{SS}$ or $\nu_{SS}$) reached during the CSR ramp. The value of the structural variable prior to the second CSR ramp is determined by the previous step in the deformation protocol, e.g., the creep step in the four-step deformation experiment. In the traditional model, the larger the creep stress (i.e., the smaller the unloading has been after the first CSR ramp) the closer $S_{in}$ is to $S_{SS}$. That is why the traditional model predicts large overshoot in case of a zero creep stress and no overshoot in case of a creep stress that is slightly below the flow stress, in contradiction to the experiment. The situation is completely reversed in the new model, where the value of $\nu_{in}$ is farther from the $\nu_{SS}$ in case of a large creep stress and closer to $\nu_{SS}$ in case of a small creep stress, resulting in larger stress overshoot with increasing creep stress in agreement with the experiment.

The reason for this behavior of the traditional structural variable $S$ is that it is inextricably and monotonically fused with mobility, which in turn is strongly correlated with deformation. During the creep step, a material under larger stress possesses higher mobility and, hence, a higher value of $S$; conversely, a material under smaller stress possesses lower mobility and, hence, a lower value of $S$. This forces $S_{in}$ to be closer to $S_{SS}$ in case of a larger creep stress with an inescapable consequence of smaller overshoot during subsequent CSR loading. In light of this monotonic relationship between the structural variable and the mobility, there is one possible scenario under which the traditional approach could possible predict the four-step deformation experiment; specifically, if for some unknown reason the mobility during the creep step was ordered in a counter-intuitive way, where material possessed a lower mobility under higher creep stress and a higher mobility under a lower creep stress. If this were the case the specific form of the eqs 2 and 3 would have to be replaced, but the approach itself may have been salvaged. The significance of the direct optical observation of the mobility reported here is that it obviates such a scenario. As we have demonstrated above, the mobility during the four-step experiment behaves in accordance with expectations, where during the creep step it is higher when the creep stress is higher. Now there is no choice but to jettison the traditional form of the constitutive models used to describe the nonlinear viscoelastic behavior of glassy polymers. In the new model, the structural variable is not fused with the mobility. The new model produces qualitatively similar stress-strain response for two substantially different assumptions about the behavior of the relaxation time $\tau$: Case 1 where $\tau$ was the same function of stress as the one governing the kinetics of the structural variable $\nu$ and Case 2 where $\tau$ was a constant, albeit one that is significantly smaller than that of the undeformed glass.

As an added benefit, the new model offers a possible resolution of a long-standing problem found in the aging behavior of glassy polymers. This problem is a need for vertical shifting when attempting to effect the time-aging time superposition of the creep compliance or stress relaxation isotherms obtained at different aging times, i.e., the glass becomes less compliant as it becomes slower. There is an extensive body of physical aging experiments, including the initial work of Struik[48] and subsequent work by the McKenna group,[49-51] where the need for vertical shifting, in addition to the standard horizontal shifting along the log(time) axis, has been documented. While the horizontal shifting can be explained by the increase in the relaxation time $\tau$ that results from aging, there is no agreed upon explanation for the vertical shifting. Obviously, the toy model described by eqs 1,4,6, 8-10 readily provides one. In the absence of large deformation, eq 6 describes the relaxation of the efficient packing fraction $\nu$ toward the equilibrium value $\bar{\nu}$. In accordance with eq 4, as $\nu$ increases so does the modulus and an increase in the modulus manifests in a vertical shift of the compliance curve as illustrated in Figure 6.

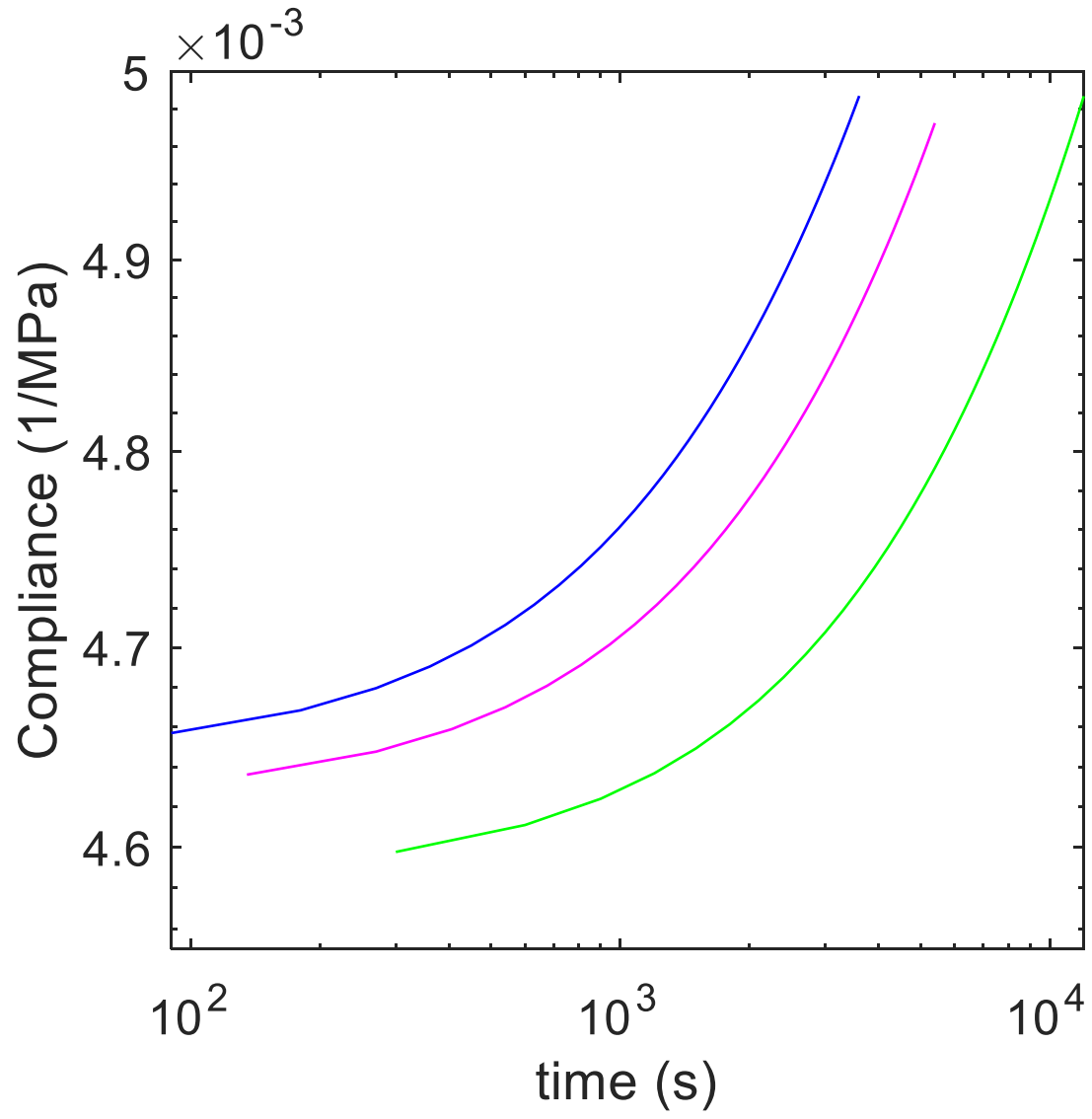

**Figure 6**. Model predictions for the Struik creep experiment – compliance vs time curves corresponding to different aging times. The model parameters are given in Table 1. The initial value for the blue curve is $\nu(0) = 0.35$ (where the equilibrium value is $\bar{\nu} = 0.71$). The initial values for the magenta and the green curves have been increased by 0.3% and 1%, respectively.

Our primary target in the current communication is not an explanation for the vertical shifting of the linear compliance and stress relaxation curves upon aging. Rather it is the nonlinear deformation experiments and specifically the second stress overshoot. So, we readily grant that alternative explanations for the vertical shifting exist. Whether these alternative models also predict the multi-step deformation experiments is a question that the authors of the models will have to contend with.

Another long-standing controversy[52] to which the model proposed here offers a possible resolution is the question whether deformation of glass results in rejuvenation or accelerated aging. As a reminder, from the perspective of the model, a material with a low value of the efficiently packed fraction $\nu$ is 'young' and a material with a high value of $\nu$ is aged; a fully equilibrated material possesses the highest value of $\nu$ (at a given temperature and in the absence of deformation). According to Eqs 6 and 8-9, applied stress affects the evolution equation for the packing fraction variable $\nu$ in two ways: first, it accelerates the overall kinetics and, second, it shifts the steady-state value lower (as breakage is favored over formation). Importantly however, these two effects do not manifest synchronously, where which is activated first depends on the values of the model parameters. This is seen in Figure 5c, where $\nu$ briefly begins to increase before turning direction and decreasing. The explanation for this behavior is as follows: at first, i.e., when the stress is still small, the rate of evolution is already high enough for $\nu$ to 'unfreeze' but the steady-state value to which it is heading is still large. Only when the stress becomes sufficiently large the target steady-state value decreases causing $\nu$ to begin decreasing as well. Thus, the model predicts that small (pre-yield) deformations result in the accelerated aging and large (post-yield) deformations result in the rejuvenation. This conclusion is in agreement with the molecular simulations of Lacks and Osborne[53] and Zhou et al.[54]

What is a possible molecular mechanism behind the model proposed here? When introducing the model, we invoked the admittedly vague idea of efficiently and inefficiently packed environments co-existing in the material, where the efficiently packed environments were assumed to have a large shear modulus and the inefficiently packed – a smaller shear modulus. The macroscopic modulus was then obtained as a weighted average given by eq 4, which implies that the environments are connected in parallel. This is an unavoidably crude attempt to describe with a single macroscopic equation what is essentially a mesoscopic phenomenon. At this point we have no experimental evidence that domains with different values of local modulus exist, although the picture of a glass forming material being a patchwork of solid-like and liquid-like domains has been a fixture in the field. A suggestive result comes from computer simulation of a glassy medium at a microscale. De Pablo and co-workers investigated the distribution of the elastic modulus at an atomic scale using a combination of MD and Monte Carlo methods,[55] where they found a normal distribution that was extremely broad, where a significant fraction of the micro-environments even had a negative modulus. Thus, in the simplest approximation one has two different moduli like the current toy model with its two discrete values of the modulus.

Finally, a comment about our perspective with respect to the two toy models. Since the point-of-departure for nonlinear viscoelastic models is the linear Maxwell model, i.e., eq 1, (or its multi-modal generalization), this is the place to expose the critical nonlinear components needed to describe the multi-step loading experiments, although there are other possible points-of-departure, e.g., the fractional calculus representation of linear relaxation.[56] The Maxwell model has just two material parameters into which to introduce nonlinearity: the relaxation time $\tau$ and the modulus $G$. The approach that underlies the current constitutive models for glassy materials is that $G$ is not affected by deformation and $\tau$ is affected either (i) directly by deformation, i.e., via $\tau(\sigma)$ or $\tau(\varepsilon)$ or $\tau(d\varepsilon/dt)$, or (ii) indirectly via $\tau(S)$, where $S$ is a structural variable that can evolve under the influence of the deformation. This was the key postulate in the first toy model, i.e., eqs 1-3, where it has been shown that although $\tau(S)$ can describe key features of a constant strain rate loading experiment, it is unable to describe the second overshoot peak (and how it depends upon the unloading stress) for the four-step experiment. The other limiting case is to consider a structural variable $S$ that affects the modulus $G$, while not affecting the relaxation time $\tau$ – this was the second toy model, i.e., eqs 1, 4, 6 and 8-10. This second toy model was able to qualitatively describe the stress strain response for the four-step deformation, even for the case when $\tau$ was a constant, although prediction of the yield stress for the constant $\tau$ case requires using a $\tau$ associated with the deformed state not the $\tau$ in the linear viscoelastic limit (see SI for a more detailed discussion). The second toy model showed that having the limiting case of $G$ depend upon a structural variable $\nu$ is able to predict the four-step experiment. Thus, we have examined the two limiting cases for a Maxwell model: $\tau(S)$ with constant $G$ that is unable to predict the four-step experiment but that can predict linear viscoelastic relaxation as well as the yield stress with post-yield softening and a second limiting case where $G(\nu)$ and $\tau$ is independent of $\nu$ that predicts the four-step deformation as well as yield with post-yield softening. The two limiting case toy models examined in this paper show the critical features of including a structural variable dependence in the two material parameters in the Maxwell model, where a profitable topic of future research would be the investigation of a model that includes two structural variables $S$ and $\nu$ with two associated evolution equations and $\tau(S)$ and $G(\nu)$ as well as including a spectrum of relaxation times – all of which may be required to quantitatively describe multi-step deformation of glassy polymers. Although the toy model described in this paper exposes the key understanding needed to describe the nonlinear mechanical behavior of glassy polymers, the development of a full constitutive description will require the use of proper finite strain and stress tensors as well as any constraints imposed by the second law of thermodynamics.

## Conclusions

A four-step deformation experiment on PMMA glass was performed while simultaneously monitoring the molecular mobility in the material using the photobleaching technique. In agreement with previous reports, the magnitude of the second stress overshoot increased as the stress level during the partially unloaded creep deformation increased, which is in contradiction to the predictions of the existing constitutive models for the deformation of glassy polymers. A representative toy model of the traditional constitutive models was able to show the reason for the discrepancy. Specifically, traditional models postulate that the stress-strain behavior of a glassy polymer is governed by the mobility (i.e., the inverse of the relaxation time) that is controlled by an internal variable describing structure of the material, where there is a monotonic relationship between the structural state of material and its molecular mobility such that a 'young' material has higher mobility than an aged one. Mechanical deformation rejuvenates the material increasing its mobility, which is supported by the optical probe measurements for both single step and the four-step deformations that showed that a partially unloaded material has higher mobility than a fully unloaded material. Since in the traditional glassy constitutive models the magnitude of the stress overshoot increases with a decrease in the mobility prior to the constant strain rate loading, the optical experiments have identified a fundamental flaw in the traditional constitutive models for glassy polymers. Using a second toy model an alternative approach has been proposed, where the effect of structural variable is contained in that the modulus vs. the relaxation time. The new model successfully describes the four-step experiment. It also explains the need for vertical shifting when describing the effect of physical aging on linear creep compliance – a long-standing problem in the field of polymeric glasses.

## Supporting Information

Evolution of the KWW parameter $\beta$ as determined from the photobleaching experiment accompanying four-step deformation shown in Figure 3; assessment of the effect of varying local strain rate on the magnitude of the second stress overshoot in the four-step deformation experiment; predictions of the four-step experiment using the Case II model.

## Acknowledgements

GAM and JMC acknowledge support from the National Science Foundation Grant Number 1761610-CMMI. EX and MDE acknowledge support from National Science Foundation (DMR) Grant Number 2002959.

# Supporting Information for "Multi-step deformation experiment and development of a model for the mechanical behavior of polymeric glasses"

Grigori A Medvedev*,[1] Enran Xing,[2] Mark D Ediger[2] and James M Caruthers[1],
Purdue University,[1] West Lafayette, Indiana 47907, USA and University of Wisconsin-Madison,[2] Madison, Wisconsin 53706, USA
*medvedev@purdue.edu

## The effect of strain rate on the second stress overshoot

When using the same global strain rate in the first and second constant strain rate steps, the local strain rate at the second yield point could be up to 3 times higher than that at the first yield point, which could potentially explain the existence of the second stress overshoot. To determine if this was the source of the second overshoot, the global strain rate was reduced from 2.1×10$^{-5}$ s$^{-1}$ to 1.2×10$^{-5}$ s$^{-1}$ in the second constant strain rate step, so that the local strain rates at both the first and the second yield points were approximately 2.5×10$^{-5}$ s$^{-1}$. The results of these experiments are shown in Fig. S1 and S2.

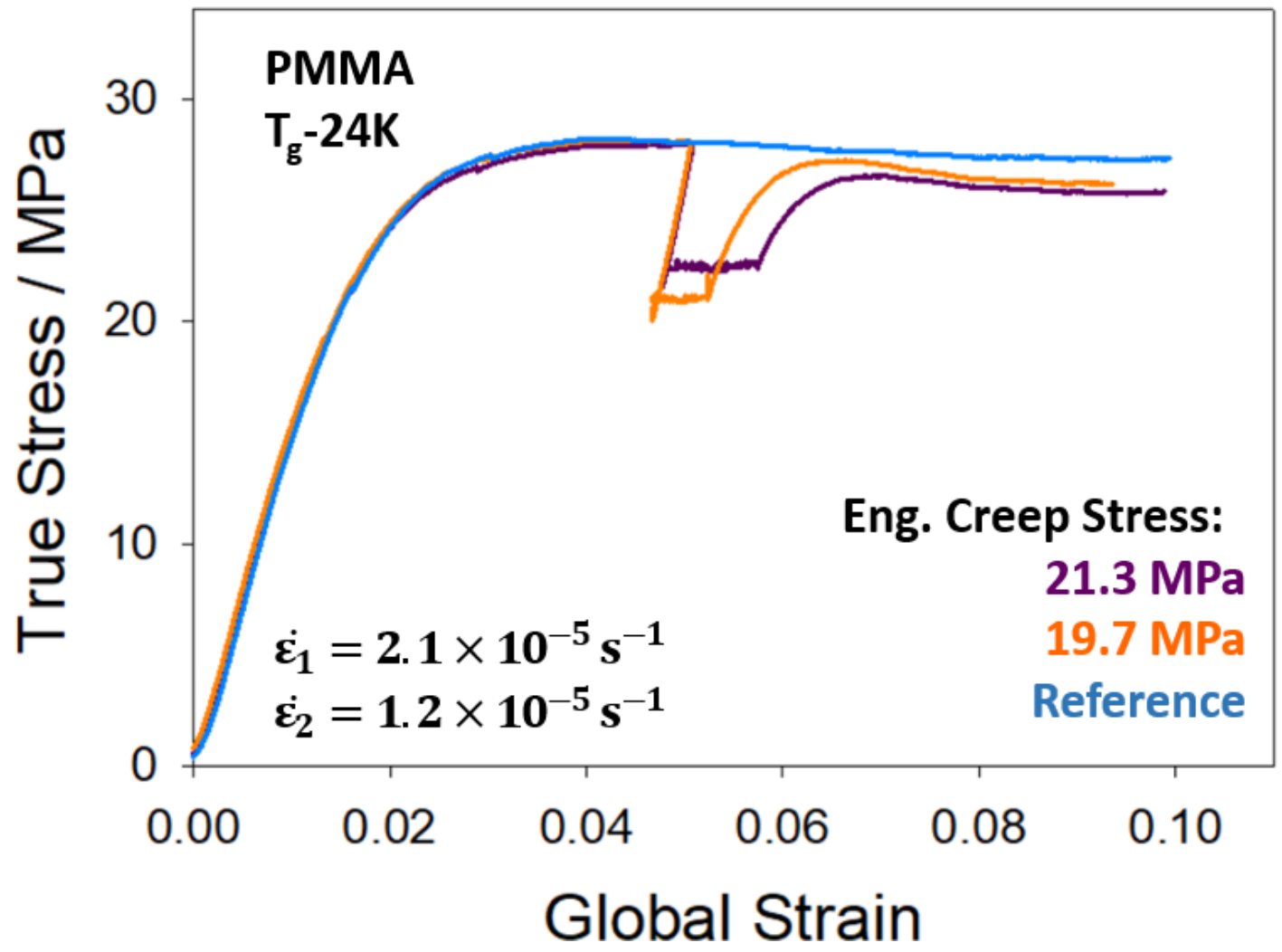

Figure S1. True stress plotted with respect to global strain for PMMA glasses deformed by multistep constant rate deformation and single-step constant strain rate deformation. The strain rate for the first step of the multistep constant strain rate deformations and the single-step constant strain rate deformation was 2.1×10$^{-5}$ s$^{-1}$, while that of the second step of the multistep constant strain rate deformations was reduced to 1.2×10$^{-5}$ s$^{-1}$. Multistep constant strain rate deformations with two creep stress levels were tested, both of which showed the second stress overshoots with a magnitude (about 1 MPa) comparable to that of the first yielding process.

In Fig. S1, multistep constant strain rate deformations with similar local strain rates at the first and second yield points show second stress overshoots with a magnitude (about 1 MPa) comparable to that of the first yielding process. Thus, the second stress overshoot shown in Fig. 3 in the main text is not a result of the increased local strain rate.

The results in Fig. S1, in the conventional view, imply a large acceleration of the segmental mobility during the second yielding process. This hypothesis was tested using the photobleaching measurements for the deformations reported in Fig. S1. In Fig. S2 the segmental dynamics are shown for the multistep constant strain rate deformations shown in Fig. S1, where a much smaller acceleration during the second constant strain rate step is observed (i.e., approximately 0.5 decade) than that during the first constant strain rate step (i.e., approximately 1.7 decades). This is consistent with data shown in Fig. 3 in the main text; thus, even when the local strain rate is approximately constant, the change of the segmental dynamics still cannot explain the appearance of the second stress overshoot.

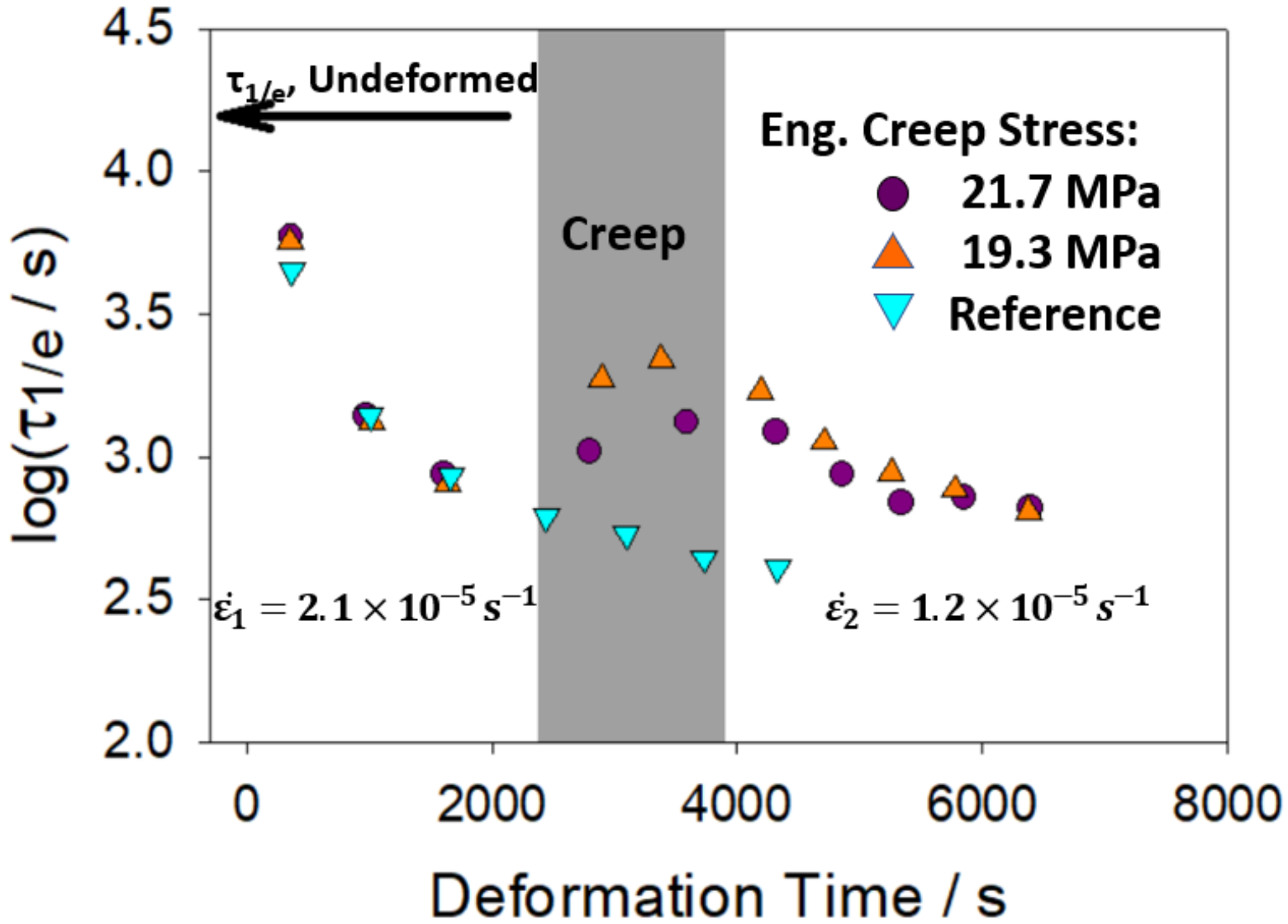

Figure S2. Evolution of the segmental mobility, $\log(\tau_{1/e})$, with respect to the deformation time for deformations shown in Fig. S1. The acceleration of the segmental dynamics during the second constant strain rate step of the multistep constant strain rate deformations was much smaller than that during the first constant strain rate step.

## Multistep constant strain rate deformations without second constant strain rate step

Control experiments were performed in which the stress was released after the creep step in a multistep constant strain rate deformation in order to determine how the segmental mobility evolves without the second constant strain rate deformation. As shown in Fig. S3 and S4, in the absence of loading the segmental relaxation time $\tau_{1/e}$ measured by the photobleaching technique gradually returns to the pre-deformation value.

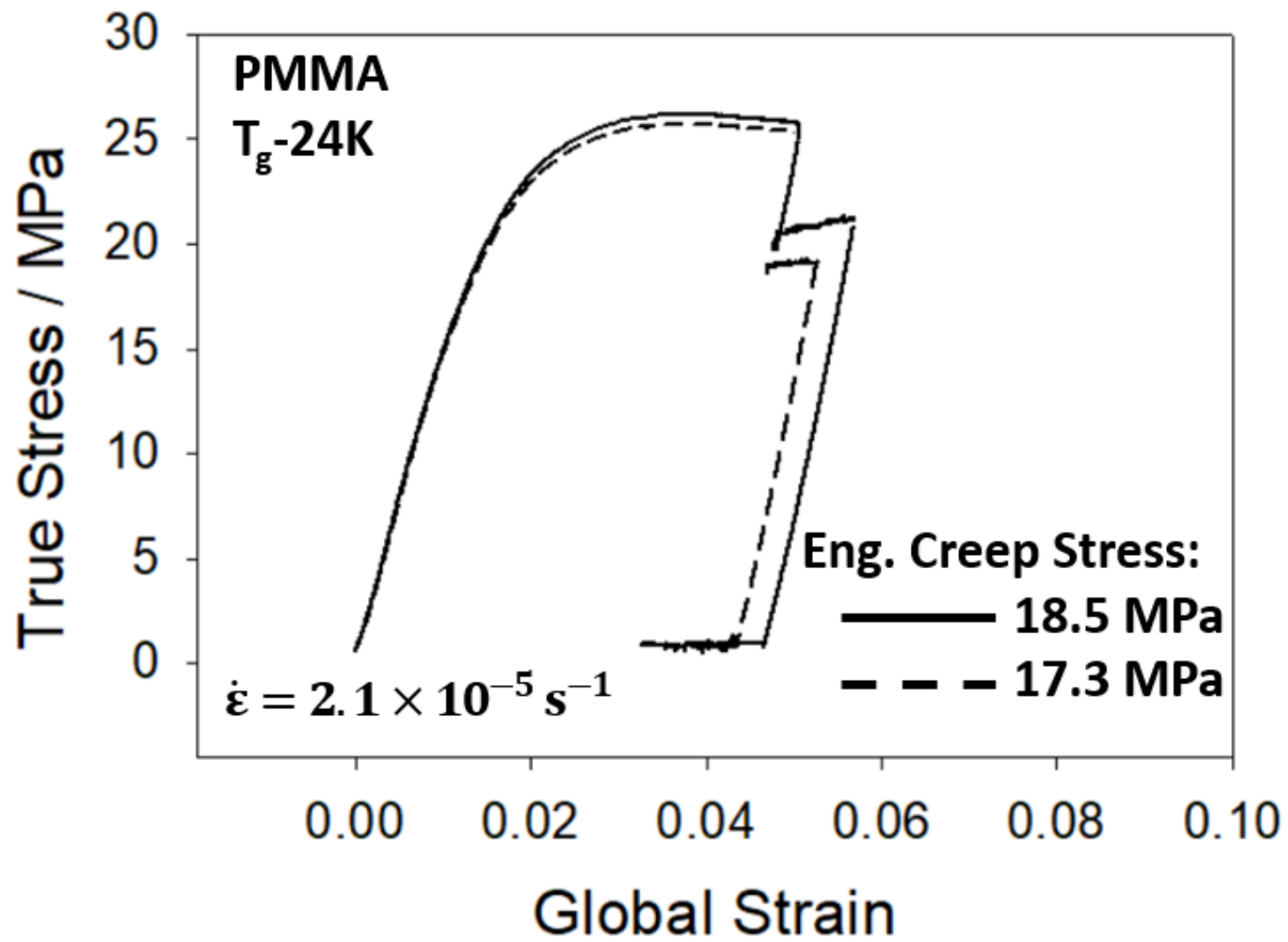

Figure S3. True stress profile with respect to global strain for deformations in which the stress was released after the creep deformation step.

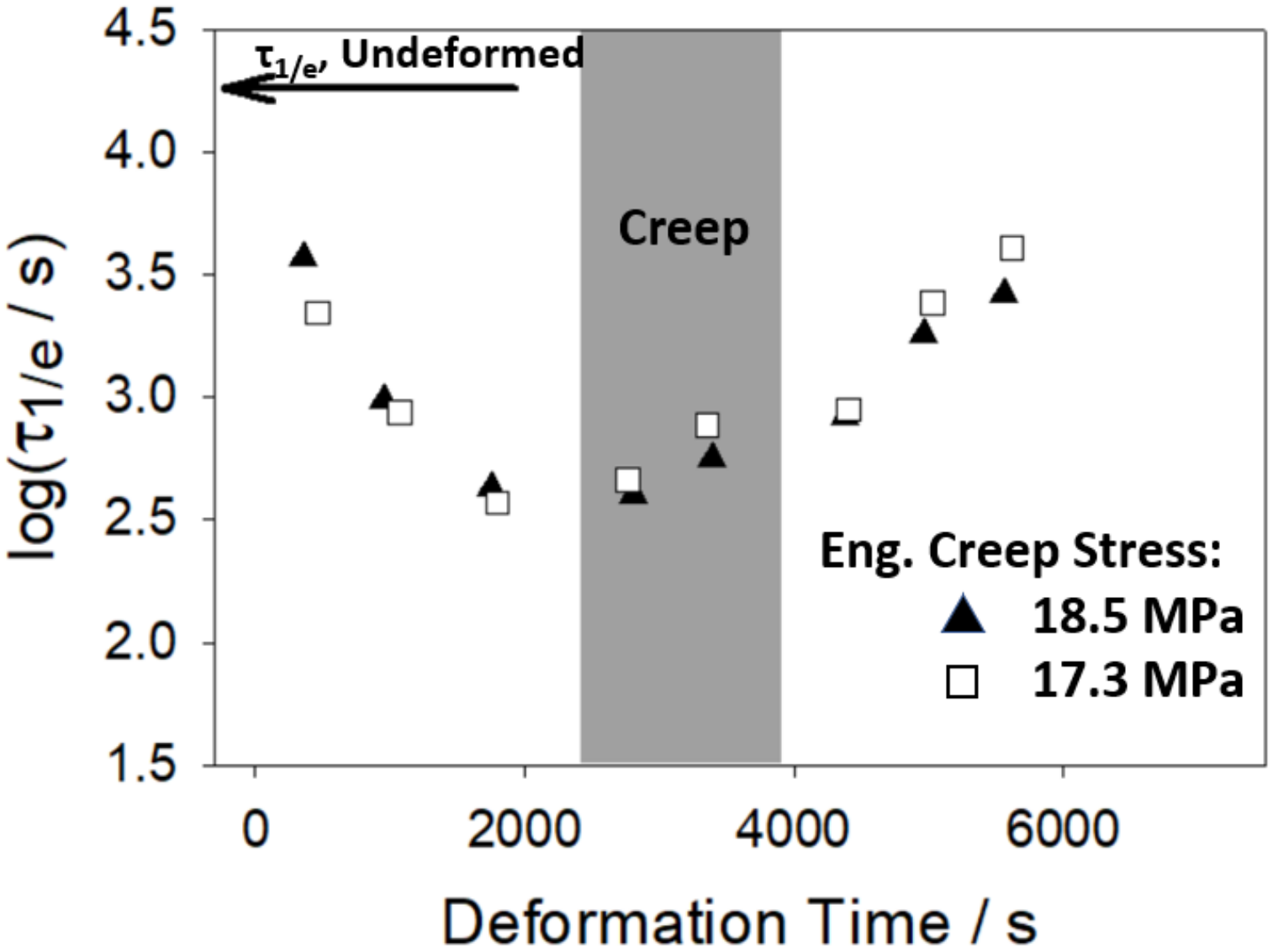

Figure S4. The evolution of the segmental dynamics for deformations shown in Fig. S3.

## Evolution of β during deformations using four-step protocol

Fig. S5 shows the evolution of $\beta$ during four-step deformations. The value of $\beta$ increased from 0.31 in the undeformed state to around 0.6 after deformations started. And during creep, $\beta$ decreased, where lower creep stress gave smaller $\beta$. However, this decrease in $\beta$ was partly caused by the deceleration of segmental dynamics, which broadened the spectrum of relaxation times over the course of one photobleaching measurement.

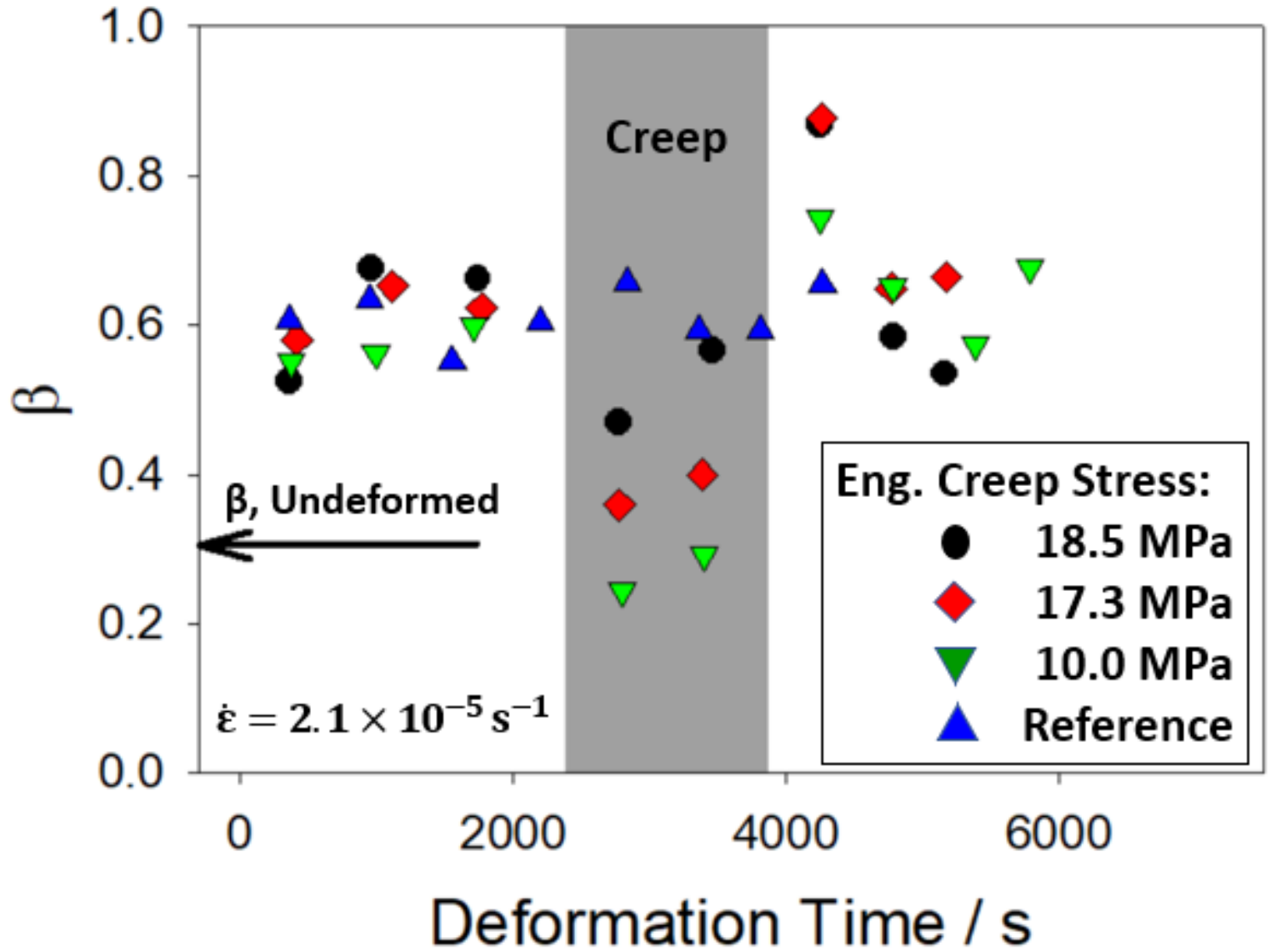

Figure S5. Evolution of $\beta$ as a function of deformation time during the deformations shown in Fig. 3 in the main text. The black arrow indicates $\beta$ value of undeformed samples. The gray shaded area shows the time period during which creep deformation was performed in the four-step deformation experiments.

## Predictions by the Toy Model in Case of Constant τ (Case II)

As stated in the main text, qualitative predictions of the toy model remain unchanged even in case of the relaxation time $\tau$ in eq 1 being kept constant. This is called the Case II in the main text, as opposed to the Case I, where $\tau$ has the same dependence on the deformation (specifically, the stress $\sigma$) as the inverse formation rate $k_f$ as given by eq 8. Predictions of the four-step experiment using the Case II model are shown in Fig S6, where the strain rate during loading and unloading steps was $10^{-3}$ $s^{-1}$. The values of the model parameters are the same as in the Case I and are given in the Table 1 in the main text.

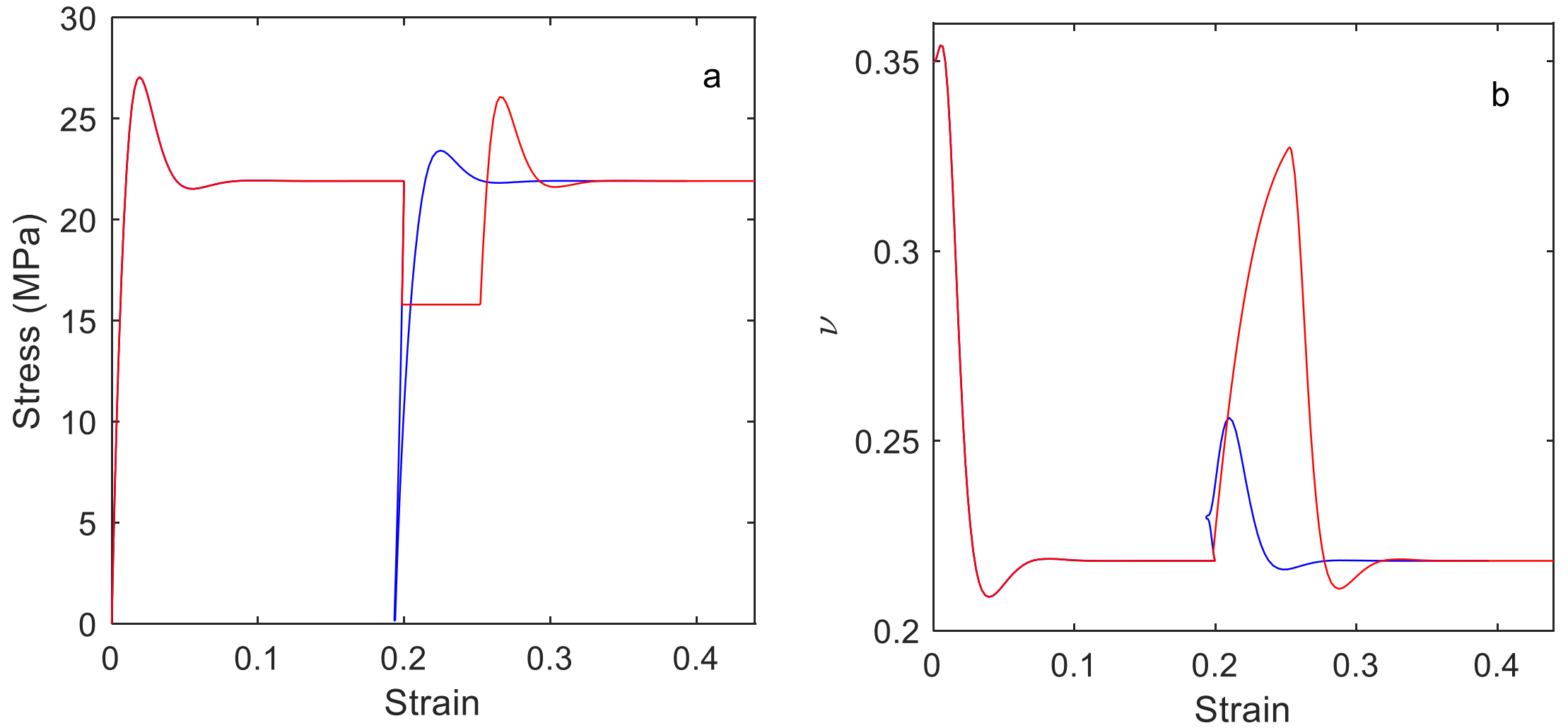

Figure S6. Multi-step experiment: (a) stress and (b) efficiently packed fraction. Unloading to zero stress – blue curve, partial unloading – red curve.

The behavior shown in Fig S6 is an almost exact copy of the behavior shown in Fig. 5 in the main text. Obviously, such a close resemblance is a result of a particular choice of the value of $\tau$, which is 10 s for the predictions in Fig. S6. Since $\tau$ is a constant, it is the same both prior to deformation and in the flow state; however, in Case I the value prior to deformation is $\tau = \tau_0 \exp(b)$ i.e., $5.4\times10^3$ s, which then decreases to 10 s in the flow state in line relaxation time measured by the optical technique for a material at approximately $T_g$-20 K. In other words, in the Case II the relaxation time $\tau$, although constant, is already set to a value it would have in the flow regime, but it is unrealistically small in the undeformed glassy state. When a constant $\tau$ is set at a larger value, for example $\tau = \tau_0 \exp(b)$, the second stress overshoot disappears. Notwithstanding the difficulties of the Case II model to describe the relaxation behavior both prior to deformation and in the flow state, Case II when $\tau$ is constant predicts the second stress overshoot.